\documentclass[lettersize,journal]{IEEEtran}
\usepackage{cite}
\usepackage{amsmath,amssymb,amsfonts,subfig}
\usepackage{algorithmic}
\usepackage{graphicx,color}
\usepackage{textcomp}
\usepackage{diagbox}
\usepackage{multirow}

\def\cA{{\mathcal{A}}}   
   \def\cH{{\mathcal{H}}}
  \def\cK{{\mathcal{K}}} \def\cL{{\mathcal{L}}}
   \def\cP{{\mathcal{P}}}
\def\cQ{{\mathcal{Q}}}

\def\diag{\mathop{\mathrm{diag}}}

\def\trace{\mathop{\mathrm{tr}}}

 \def\bphi{{\pmb{\phi}}}

\def\b0{{\pmb{0}}} 

\def\ba{{\mathbf{a}}} \def\bb{{\mathbf{b}}}  \def\bd{{\mathbf{d}}}
 \def\bff{{\mathbf{f}}}  \def\bh{{\mathbf{h}}}
   
 \def\bn{{\mathbf{n}}}  
   
 \def\bv{{\mathbf{v}}}

\def\bA{{\mathbf{A}}}   
 \def\bF{{\mathbf{F}}}  \def\bH{{\mathbf{H}}}
\def\bI{{\mathbf{I}}}   
   \def\bP{{\mathbf{P}}}

\newcolumntype{b}{X}
\newcolumntype{s}{>{\hsize=.60\hsize}
	>{\centering\arraybackslash}X}
\def\BibTeX{{\rm B\kern-.05em{\sc i\kern-.025em b}\kern-.08em
    T\kern-.1667em\lower.7ex\hbox{E}\kern-.125emX}}
\usepackage{balance}
\begin{document}
\title{WiThRay: A Versatile Ray-Tracing Simulator for Smart Wireless Environments}
\author{Hyuckjin Choi\thanks{H. Choi and J. Choi are with the school of Electrical Engineering, Korea Advanced Institute of Science and Technology (e-mail: \{hugzin008,junil\}@kaist.ac.kr).}, Jaehoon Chung, Jaeky Oh\thanks{J. Chung and J. Oh are with 6G Core R\&D Technology Platform (TP), Communication \& Media Standard Lab., ICT Tech. center, CTO Division, LG Electronics, Inc. (e-mail: \{jaehoon.chung,jaeky.oh\}@lge.com).}, George C. Alexandropoulos\thanks{G. C. Alexandropoulos is with the Department of Informatics and Telecommunications, National and Kapodistrian University of Athens, Panepistimiopolis Ilissia, 15784 Athens, Greece (e-mail: alexandg@di.uoa.gr).}, and Junil Choi}

\maketitle

\begin{abstract}
This paper presents the development and evaluation of WiThRay, a new wireless three-dimensional ray-tracing (RT) simulator. RT-based simulators are widely used for generating realistic channel data by combining RT methodology to get signal trajectories and electromagnetic (EM) equations, resulting in generalized channel impulse responses (CIRs). This paper first provides a comprehensive comparison on methodologies of existing RT-based simulators. We then introduce WiThRay, which can evaluate the performance of various wireless communication techniques such as channel estimation/tracking, beamforming, and localization in realistic EM wave propagation. WiThRay implements its own RT methodology, the bypassing on edge (BE) algorithm, that follows the Fermat's principle and has low computational complexity. The scattering ray calibration in WiThRay also provides a precise solution in the analysis of EM propagation. Different from most of the previous RT-based simulators, WiThRay incorporates reconfigurable intelligent surfaces (RIS), which will be a key component of future wireless communications. We thoroughly show that the channel data from WiThRay match sufficiently well with the fundamental theory of wireless channels. The virtue of WiThRay lies in its feature of not making any assumption about the channel, like being slow/fast fading or frequency selective. A realistic wireless environment, which can be conveniently simulated via WiThRay, naturally defines the physical properties of the wireless channels. WiThRay is open to the public, and anyone can exploit this versatile simulator to develop and test their communications and signal processing techniques.
\end{abstract}


\section{Introduction}
\label{sec:introduction}
\IEEEPARstart{M}{ost} wireless communication research is based on channel models. Among many possible models, the geometry-based stochastic channel model (GSCM) is widely used in both academia and industry \cite{meinila2009winner,Lin:Oestges:COST2100,Ghazal:2017,series2017guidelines}. Several technical reports provide guidelines for designing the GSCM, where intensive measurement campaigns for particular scenarios, e.g., urban micro cellular (UMi), urban macro cellular (UMa), and indoor, have determined the channel parameters \cite{3gpp2017study,3gpp2019study,series2017guidelines}.

The latest wireless communication techniques have largely exploited the geometrical properties of the radio frequency (RF) channel at high-frequency bands such as millimeter-wave (mmWave), and very recently terahertz (THz). The geometrical sparsity of the mmWave channel forces the rank deficiency for the high-dimensional channel, making highly complicated systems tractable \cite{Alkhateeb:2014_2,Alkhateeb:2015,Sun:2017}. The assumption about channel sparsity enables the compressive sensing (CS)-based approaches on hybrid beamforming systems\cite{Xinying:2005,Alkhateeb:2014_1,Molisch:2017, Vlachos:2019}, reconfigurable intelligent surface (RIS)\cite{Wang:2020,He:2020,Wei:2021,Wei:2022,Hong:2022}, and cell-free multiple-input and multiple-output (MIMO) systems \cite{Femenias:2019}.

The channel sparsity is evident on the angular domain. The angular spread profile, a picture of the RF signals coming from various directions, confines the covariance matrix of the GSCM to be low-rank. The GSCM relies on stochastic angular spread profiles given with random cluster positions and trajectories of channel paths. The power profiles on the temporal and spectral domains are also stochastic parameters for the GSCM \cite{series2017guidelines}. If we are able to obtain site-specific, not stochastic, angular spread profiles, there would be room to further improve wireless communication techniques by site-specifically optimizing them \cite{He:2019}.

Without intensive measurement, the ray-tracing (RT) algorithm gives the detailed geometrical relationship between communication terminals in a deterministic manner \cite{Yun:2015,He:2019}. This deterministic formalism then employs the electromagnetic (EM) equations, which have no randomness when the geometrical parameters are entirely given. There exist various wireless channel simulators using the RT algorithm in academia \cite{Tan:1996,Son:1999,Adana:2005,Boban:2014,Fuschini:2015,Wang:2016,Lecci:2020,Choi:2021} and industry\cite{Hoydis:Sionna,Chen:2021,Sihlbom:2022,WinProp,WirelessInSite,Ranplan,EDX,Volcano,CrossWave,Aster}. Commercial programs have their own acceleration algorithms to boost up the RT algorithm such that they can conduct large-scale simulations; however, the details of simulators are not open to public in general.

The RT algorithm is to find the channel paths following the Fermat's principle. The methods to implement the RT algorithm are largely divided into two types; the direct algorithm and the inverse algorithm \cite{Catedra:1998}. The direct algorithm tracks all rays that propagate omnidirectionally. Each ray changes the propagating direction when meeting obstacles, i.e., reflections or diffractions. The direct algorithm is also called as the ray launching (RL) method or the shooting and bouncing ray (SBR) method in related works \cite{Fuschini:2015}.

The inverse algorithm aims to determine the accurate channel path trajectories from the transmitter (Tx) to the receiver (Rx). The inverse algorithm provides more precise simulation results than the direct algorithm by considering all possible connectivity. However, with an increasing number of objects in the simulation, the complexity of the inverse algorithm becomes challenging. To mitigate the complexity of inverse algorithm, various methods were proposed, e.g., the visible-tree (VT) method and grid partitioning methods such as binary space partitioning (BSP) tree, space volumetric partitioning (VSP) method, and angular Z-Buffer (AZB) method\cite{glassner1989into,Sanchez:1996,Catedra:1998}.

\begin{table}[t]
	\centering
	\caption{Notations used in this paper.}
	\begin{tabular}{|l|l|}
		\hline
		$\ba$ & vector \\ \hline
		$\bA$ & matrix \\ \hline
		$\boldsymbol{\cA}$ & tensor \\ \hline
		$\b0_{p,q}$ & $p\times q$ matrix with zero elements \\ \hline
		$\bA^{-1}$ & inverse of matrix $\bA$ \\ \hline
		$\bA^\mathrm{T}$ & transpose of matrix $\bA$ \\ \hline
		\multirow{2}{*}{$\bA^\mathrm{H}$} & conjugate transpose (Hermitian) of \\
		& matrix $\bA$ \\ \hline
		$|a|$ & absolute value of $a$ \\ \hline
		$\lVert\ba\rVert$ & $\ell_2$ norm of vector $\ba$\\ \hline
		tr\{$\bA$\} & trace of matrix $\bA$ \\ \hline
		\multirow{2}{*}{diag\{$\ba$\}} & diagonal matrix with $\ba$ as diagonal \\
		& elements \\ \hline
		$[\ba]_i$ & $i$-th element of vector $\ba$ \\ \hline
		$[\bA]_{i,j}$ & $(i,j)$-th element of matrix $\bA$ \\ \hline
		\multirow{2}{*}{$[\bA]_{i:i+k,j:j+\ell}$} & sub-matrix with $i$-th to $(i+k)$-th rows \\
		& and $j$-th to $(j+\ell)$-th columns of $\bA$ \\ \hline
		$\ba\cdot\bb$ & inner product of vectors $\ba$ and $\bb$ \\ \hline
		$\ba\times\bb$ & outer product of vectors $\ba$ and $\bb$ \\ \hline
		$f(t)\ast g(t)$ & convolution of $f(t)$ and $g(t)$ \\ \hline
	\end{tabular}
\end{table}

Reflection, diffraction, and scattering are key factors in the implementation of a high-performance RT-based simulator. When considering the scattering, it should be noted that the reflecting and diffracting paths do not always move on the specular channel path trajectory. Thus, to evaluate the channel impulse responses (CIRs) under the scattering, an appropriate definition for the scattering channel paths and EM response models should be given. Most of existing simulators, however, do not implement all these factors simultaneously \cite{Tan:1996,Son:1999,Adana:2005,Boban:2014,Lecci:2020}.

In this paper, we present a versatile wireless channel simulator, named as WiThRay (Wireless three-dimensional ray-tracing simulator). Most of RT-based simulators developed for wireless communication systems focus on the EM field examination \cite{Adana:2000,Guan:2013,Degli:2014,Chang:2014,Hur:2016,Lecci:2021}. However, WiThRay is specialized to generate the channel data for performance evaluations of complicated communication techniques. WiThRay gives the channel data with fundamental channel characteristics, e.g., short coherence time for large Doppler spread and small coherence bandwidth for large delay spread. The WiThRay channel data are saved into two types; multi-tap channels in the time domain and subcarrier channels in the frequency domain, making WiThRay suitable to evaluate most of wireless communication techniques. The main features of WiThRay are summarized below:

\begin{enumerate}	
	\item[i)] WiThRay employs its own RT algorithm, dubbed as the bypassing on edge (BE) algorithm, which enhances the efficiency of the inverse algorithm.
	\item[ii)] WiThRay dynamically adjusts the scattering area to include the dominant channel paths.
	\item[iii)] WiThRay calibrates the scattering grid for MIMO scenarios to ensure that the EM wave propagates in its original waveform, as it does in the real world.
	\item[iv)] WiThRay judiciously evaluates the multi-tap channel in the fast fading scenario for the development of high-mobility communication systems. The wideband effect and Doppler shift caused by mobile objects are modeled in WiThRay.
	\item[v)] WiThRay supports the RIS systems by employing the dyadic EM response model.
\end{enumerate}

WiThRay is open to public in our website \cite{WiThRay}, and anyone can use this versatile simulator to evaluate various wireless communication techniques. The website \cite{WiThRay} also contains the implementation details of WiThRay.

The rest of this paper is organized as follows. Section \ref{sec2} provides the baseline of the general RT algorithm and compares the existing RT-based simulators. Section \ref{sec3} details the technical features of WiThRay, including the BE algorithm, scattering calibration, geometrical parameter evaluation, and CIR modeling. The output of WiThRay is the discrete channel data, and the discrete sampling process is also described in Section \ref{sec3}. Extensive experimental results are provided in Section \ref{sec4} showcasing that the channel data generated by WiThRay follow the fundamental theory of wireless channels. After analyzing several wireless communication techniques with WiThRay in Section \ref{sec5}, concluding remarks of the paper are included in Section \ref{sec6}.

\begin{table*}[t]
	\centering
	\caption{Implementation details of RT algorithms.}
	\begin{tabular}{|c|c|c|c|c|c|}
		\hline
		\multirow{2}{*}{\textbf{Simulators}} & \multicolumn{5}{c|}{\textbf{RT algorithm}} \\ \cline{2-6} 
		& \textbf{methodology} & \textbf{reflection} & \textbf{diffraction} & \textbf{scattering} & \textbf{quasi-3D/full-3D} \\ \hline 
		\multirow{2}{*}{WiThRay (proposed)} & \multirow{2}{*}{inverse (BE)}  & \multirow{2}{*}{multiple} & \multirow{2}{*}{multiple} & adjustable scattering area, & \multirow{2}{*}{full-3D} \\
		&  & & & MIMO calibration &  \\ \hline
		Ray tube \cite{Tan:1996,Son:1999} & inverse (VT \cite{Sanchez:1996}) & multiple & multiple & - & quasi-3D \\ \hline
		FASPRO \cite{Adana:2005} & inverse (AZB \cite{Catedra:1998}) & double & single & - & full 3D \\ \hline
		GEMV \cite{Boban:2014} & inverse & single & multiple & - & full 3D \\ \hline
		\multirow{2}{*}{3DScat \cite{Fuschini:2015,Degli:2004,Degli:2007,Bilibashi:2020}} & direct, & \multirow{2}{*}{multiple} & \multirow{2}{*}{multiple} & \multirow{2}{*}{Fresnel's ellipsoid \cite{Degli:2004}} & \multirow{2}{*}{full 3D} \\
		& inverse (VT \cite{Sanchez:1996}) & & & & \\ \hline
		CloudRT \cite{Wang:2016, Guan:2013} & inverse \cite{McKown:1991} & multiple & - & - & full 3D \\ \hline
		Simplified ray tracer \cite{Lecci:2020} & inverse & multiple & - & - & full 3D \\ \hline
		\multirow{2}{*}{Wireless InSite \cite{WirelessInSite}} & direct (SBR \cite{Ling:1989}), & 30 (SBR), & 4 (SBR), & \multirow{2}{*}{elliptical region} & \multirow{2}{*}{full 3D} \\ 
		& inverse (Eigen ray) & 3 (Eigen ray) & 4 (Eigen ray) & & \\ \hline
	\end{tabular}
	\label{table:RTalg}
\end{table*}

\begin{figure}
	\centering
	\includegraphics[width=0.88\columnwidth]{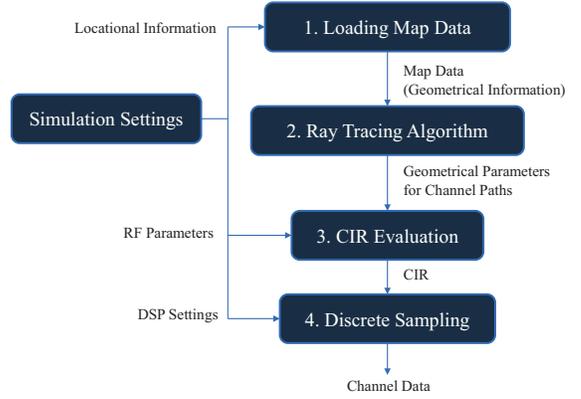}
	\caption{Flowchart of general RT-based simulator.}
	\label{fig:1}
\end{figure}

\section{RT-based Simulators}
\label{sec2}
An RT-based simulator is a computer tool used to model the EM propagation in a three-dimensional (3D) environment. The RT algorithm traces EM waves as they propagate through the environment. The simulation considers various factors, such as reflection, diffraction, and scattering, which can affect the signal quality in a real-world environment. The simulation results can be used to predict the signal quality, data rate, and coverage area of a wireless communication system, providing crucial information for the design and optimization of the system.

The general procedure of most RT-based simulators is shown in Fig.~\ref{fig:1}. A simulator should first build a realistic environment that determines the trajectories of RF signals from the Tx to the Rx. The RT algorithm finds the geometrical channel paths and provides geometrical parameters, e.g., propagation direction, polarization direction, traveling distance, delay, angle of departure, and angle of arrival, which are required to evaluate the CIR. The evaluated CIR should be sampled to test various digital signal processing (DSP) techniques. The details of the procedure are presented in the following:

\subsection{Loading Map Data}
The RT-based simulators generate wireless channel data that change based on the surrounding environment. The RT-based simulators consider important map components that define the channel. In outdoor environments, the dominant components in the map are buildings, terrain, and in some cases, foliage. Indoor environments are largely affected by walls, doors, windows, and large objects, e.g. furniture. Communication terminals should also be loaded at this stage.

Commercial programs, such as Wireless InSite, provide their own 3D map rendering program, but they can also import external rendered data. WiThRay also composes maps by importing 3D coordinate data of objects.

The size of map used for simulation varies depending on its purpose. High-frequency communication scenarios typically assume small areas for simulations. However, for high-speed mobile communications or resolving resource allocation issues in multiple access, the map size may need to increase. Without an efficient algorithm, the RT-based simulator cannot conduct large-scale simulations. WiThRay uses a fast RT algorithm to enable large-scale simulations, which is elaborated in Section~\ref{sec3_2}.

\subsection{RT Algorithm}
The RT algorithm is a widely-used technique for simulating EM wave propagation and creating realistic computer graphics. The algorithm tracks the multiple channel paths from the Tx to the Rx, following the Fermat's principle at points of reflection and diffraction. The direct and inverse algorithms are the two key techniques used to determine these points. In Table \ref{table:RTalg}, we compare the RT algorithm adopted in WiThRay to those used in existing RT-based simulators.

\subsubsection{Methodology}
The direct algorithm shoots rays in all directions and tracks their trajectories. When these rays reach an object, they may reflect or diffract. Only the rays that reach the Rx node are transformed into useful information. The accuracy of the direct algorithm depends on the number of rays generated at each branching point (i.e., reflecting or diffracting points). The direct algorithm provides simultaneous results for multiple Rx nodes and is therefore preferable for large-scale simulations. Simulators such as 3DScat and Wireless InSite implement the direct algorithm to support city-scale simulations \cite{Fuschini:2015,WirelessInSite}. In the process of tracing multiple rays, a graphical processing unit (GPU) enables parallel operations, making computation efficient on hardware \cite{Epstein:2010}. However, the number of rays to be tracked in the direct algorithm could be redundant for the simulations with a small number of nodes. Moreover, the direct algorithm suffers from quantization errors, i.e., discretized rays satisfying the Snell's law do not precisely converge at the Rx node. Therefore, the direct algorithm relies on the concept of acceptance region around the Rx node, which cause inaccurate propagation paths.

The inverse algorithm that only finds the rays of the Fermat's principle is preferred in simulations that demand high accuracy. Several RT-based simulators, including WiThRay, are based on the inverse algorithm \cite{Tan:1996,Son:1999,Adana:2005,Boban:2014,Fuschini:2015,Wang:2016,Lecci:2020,Choi:2021,WirelessInSite}. However, the inverse algorithm can become complicated when dealing with a large number of objects in the simulation. As the number of objects increases, the number of cases need to be checked for the inverse algorithm grows exponentially. Moreover, the inverse algorithm must search for channel paths for every node-to-node connection, resulting in high complexity.

\begin{table*}[t]
	\centering
	\caption{Implementation details of CIR modelings.}
	\begin{tabular}{|c|c|c|c|c|}
		\hline
		\multirow{2}{*}{\textbf{Simulators}} & \multicolumn{4}{c|}{\textbf{CIR}} \\ \cline{2-5} 
		& \textbf{reflection} & \textbf{diffraction} & \textbf{Doppler} & \textbf{RIS} \\ \hline 
		WiThRay (proposed) & directive scattering \cite{Degli:2007} & UTD \cite{Kouyoumijan:1974} & mUE, mSO \cite{Bilibashi:2020} & Faraday's law\cite{Najafi:2021} \\ \hline
		Ray tube \cite{Tan:1996,Son:1999} & dyadic reflection \cite{stratton2007electromagnetic} & UTD \cite{Kouyoumijan:1974} & mUE & - \\ \hline
		FASPRO \cite{Adana:2005} & dyadic reflection \cite{stratton2007electromagnetic} & UTD \cite{Kouyoumijan:1974} & - & - \\ \hline
		GEMV \cite{Boban:2014} & dyadic reflection \cite{Landron:1996} & ITU-R(P)\cite{series2013recommendation} & mUE, mSO \cite{Bilibashi:2020} & - \\ \hline
		3DScat \cite{Fuschini:2015,Degli:2004,Degli:2007,Bilibashi:2020} & directive scattering \cite{Degli:2007} & UTD \cite{Kouyoumijan:1974} & mUE, mSO \cite{Bilibashi:2020} & - \\ \hline
		CloudRT \cite{Wang:2016, Guan:2013} & directive scattering \cite{Degli:2007} & - & mUE, mSO \cite{Bilibashi:2020} & - \\ \hline
		Simplified ray tracer \cite{Lecci:2020} & dyadic reflection \cite{stratton2007electromagnetic} & - & - & - \\ \hline
		Wireless InSite \cite{WirelessInSite} & directive scattering \cite{Degli:2007} & UTD \cite{Kouyoumijan:1974} & mUE & support \\ \hline
	\end{tabular}
	\label{table:CIR}
\end{table*}

\subsubsection{Reflection and diffraction}
The direct algorithm determines the next direction of the ray when it encounters an object. The ray could reflect off the surface or diffract at the edge. In the direct algorithm, it is straightforward to find the reflecting path while this is not the case for the diffracting path. The Keller's cone is related to diffraction, and all directions within the cone satisfy the Fermat's principle, ensuring the shortest path \cite{Keller:62}. Because the direct algorithm cannot determine in advance which direction will lead to the Rx node, it must track all directions, resulting in an increased number of rays to be tracked at diffracting edges. However, in the direct algorithm, the calibration of reflecting and diffracting points can be done sequentially, which reduces the complexity of the process.

The inverse algorithm generates a ray by connecting the mirror image of the Tx or Rx nodes with a straight line. This approach is called as the image method. When a large number of reflections encounter numerous objects, the inverse algorithm needs to produce a significant amount of mirrored images. To confine the number of mirror images, the VT method defines the visible region, known as a ray tube \cite{Sanchez:1996}. In the VT method, the mirror images of the terminal nodes have their own ray tubes. The visibility of ray tubes restricts the number of objects to be checked, and the VT method connects nodes (as a branch of the tree) that see each other through their ray tubes.

The diffracting points in the inverse algorithm should be connected with the mirror images in the flipped space. To obtain the flipped diffraction points on the Keller's cone, the edges where the diffraction occurs should be flipped multiple times according to the number of reflections in the channel path. However, determining the exact diffracting points is a non-deterministic polynomial-time (NP)-hard problem for a large number of diffractions. WiThRay iteratively solves the distance-minimizing problem to find multiple diffracting points where the shortest channel path lies on.

The grid partitioning method is an acceleration technique that limits the objects under the diffraction. The entire map is divided into grids based on a considered metric, e.g., the volumetric cube in the VSP and the angles in the AZB. The grids under the trajectory of channel path determine possible objects that may have diffracting points. However, the grid partitioning methods might miss some possible connectivity, resulting in degraded accuracy. The BE algorithm of WiThRay only takes the valid diffracting paths in the shadowed area, which efficiently decreases the cases to be checked and achieves good accuracy. The detail of BE algorithm is explained in Section~\ref{sec3_2}.

\subsubsection{Scattering}
The paths that satisfy the Fermat's principle are called as specular paths, which can be found using direct or inverse algorithms. However, in reality, the channel paths that do not follow the Fermat's principle, known as anomalous paths, can largely affect the channel. For the single point of observation, it can be interpreted as a scattering. These paths can be constructed based on any connection of arbitrary points on objects. Thus, the effect of scattering can be evaluated by considering all the connectivity of discretized points on objects, which is referred to as the propagation graph (PG) theory \cite{Tian:2016,Chen:2017}.

The PG-based method that incorporates either the direct or inverse algorithms can focus only on effective channel paths, resulting in a reduced number of rays to be considered. Related works select anomalous paths based on path length, which can be interpreted as the paths within an elliptical region \cite{Degli:2004,Zhang:2022}. However, in Section~\ref{sec3_2}, we show that an ellipsoid with a fixed radius may not be a good criterion for choosing the anomalous paths.

\subsection{CIR Evaluation}
The RT algorithm generates the rays that provide geometrical parameters such as polarization direction, propagation direction, traveling distance, delay, angle of departure, and angle of arrival. The RT-based simulators use these parameters to evaluate the CIR for each channel path by applying them to relevant equations. The CIR is a result of antenna radiation, EM propagation, and EM responses of reflections and diffractions. Each of these factors independently affects the channel in general CIR modeling \cite{Schaubach:1992,Tan:1996,Son:1999}. In WiThRay, the RIS response is also combined at this stage. Table~\ref{table:CIR} summarizes the factors that RT-based simulators consider for the CIR evaluation.

\subsubsection{Antenna radiation pattern and propagation loss}
The propagation of EM wave from the Tx antenna is determined by the propagating direction, orientation of antenna, and antenna radiation pattern. The intensity of EM wave decreases with the traveling distance, which is called as the propagation loss. The law of energy conservation just concerns the traveling distance, but the atmosphere causes the energy loss and makes the propagation loss to be a function of the carrier frequency. Most of RT-based simulators including WiThRay consider the atmospheric absorption.

\subsubsection{Reflection}
RT-based simulators utilize either the dyadic reflection model \cite{stratton2007electromagnetic} or the directive scattering model \cite{Degli:2007} as the reflection response model. The dyadic reflection model, based on the Faraday's law, is the canonical formulation. Any EM wave has perpendicular and parallel modes that react independently on the perfectly magnetically conducting (PMC) medium. However, it only deals with the specular paths on the infinite PMC plane and cannot provide a solution for anomalous paths, which ignores the scattering modeling. Therefore, the RT-based simulators that use the dyadic reflection model in Table~\ref{table:CIR} do not support scattering \cite{Tan:1996,Son:1999,Adana:2005,Boban:2014,Wang:2016,Lecci:2020}.

Scattering models are data-fitting functions. Initially, the Lambertian scattering model provided a good approximation \cite{Degli:2001}. After the measurement campaign for the reflecting signal on buildings in \cite{Degli:2007}, the directive scattering model has become a popular model for RT-based simulators\cite{Adana:2005,Boban:2014,Fuschini:2015,Wang:2016,Choi:2021}. The directive scattering model constructs a scattering lobe towards the specular reflection, and the controllable lobe width in the model determines the degree of scattering \cite{Degli:2007}. The directive scattering model also considers the polarization effect, but it does not provide a specific model for the ray with a certain polarization direction \cite{Degli:2011}. On the contrary, WiThRay incorporates the effect of polarization into the directive scattering model.

\begin{figure}
	\centering
	\includegraphics[width=0.6\columnwidth]{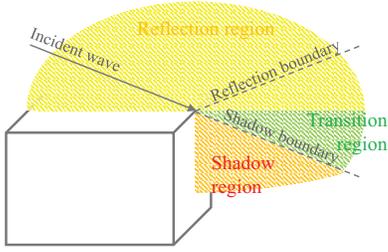}
	\caption{Incident wave at edge can diffract to reflection, transition, and shadow regions. Reflection and shadow boundaries are the directions of dominant diffraction, when the object is PMC.}
	\label{fig:2}
\end{figure}

\subsubsection{Diffraction}
The geometrical theory of diffraction (GTD) is the fundamental theory of diffraction for the RT algorithm. The GTD enables the evaluation of diffraction using rays \cite{Keller:62}, but it is an asymptotic equation that fails to solve the equation in the region close to the shadow boundary and reflection boundary in Fig.~\ref{fig:2}. The uniform theory of diffraction (UTD) provides a valid solution for the unsolved region of the GTD and generalizes the diffraction coefficients for curved wedges \cite{Kouyoumijan:1974}. Although the GTD and UTD are dyadic response models, they provide good approximations for the scattering on edges. Most of RT-based simulators that support the diffraction are based on the UTD as in Table~\ref{table:CIR}.

\subsubsection{Doppler shift}
The Doppler shift is caused by the movement of a terminal and the dynamic environment. Many RT-based simulators support the mobile user equipment (mUE) and evaluate the Doppler shift based on its relative speed to a fixed base station (BS). However, not all RT-based simulators consider mobile scattering objects (mSOs) that may affect the trajectory of channel. Without considering mSO, the range of Doppler shift is limited to the speed of mUE, but the Doppler spread might be greater than the Doppler shift by mUE, which fails to reflect real-world scenarios that have mSOs. WiThRay generates realistic channel data by considering the Doppler spread caused by the mSO.

\subsubsection{RIS response model}
The RIS is the strong candidates for 6G wireless communication systems \cite{Huang:2019,Huang:2020,Wu:2020,Jian:2022,Faqiri:2022}. In order to support an RIS system in the RT-based simulator, the EM response on the RIS panel should be evaluated in a series of reflections and diffractions. The RIS panel is a passive element that adjusts the reflections to the intended direction. The reflection of the RIS on a PMC surface is derived using the EM equivalent theorem \cite{Najafi:2021}, which is adopted by WiThRay.

\subsection{Discrete Sampling}
The discrete channel data at the BS are observable and controllable signals that can be processed at the DSP unit of wireless communication systems. However, many RT-based simulators do not provide the channel data in the discrete domain e.g., since Wireless InSite does not provide discrete-time domain channel data, DeepMIMO converts the CIR data obtained using Wireless InSite into the discrete channel data \cite{Alkhateeb:DeepMIMO}. On the contrary, WiThRay not only provides the CIR data but also the discrete channel data simultaneously, making the simulator much more efficient. The detailed process of obtaining the discrete channel data is explained in Section~\ref{sec3_6}.

\begin{figure}
	\centering
	\begin{center}$
		\begin{array}{c}
			\subfloat[Map data imported from the commercial website.]{\includegraphics[width=0.92\columnwidth]{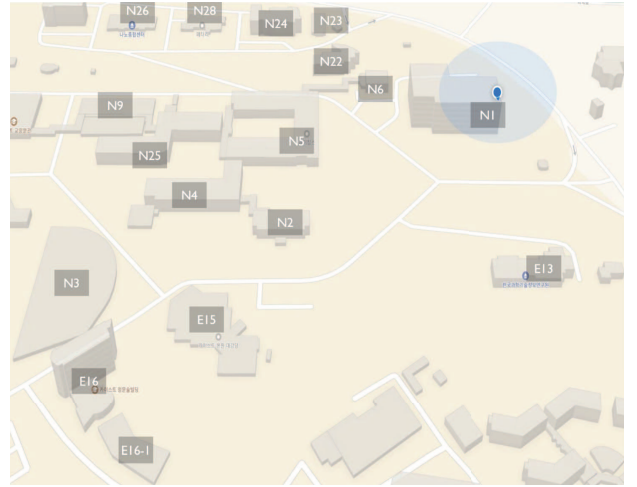}} \label{subfig:3-a}\\
			\subfloat[3D map rendered in WiThRay.]{\includegraphics[width=0.92\columnwidth]{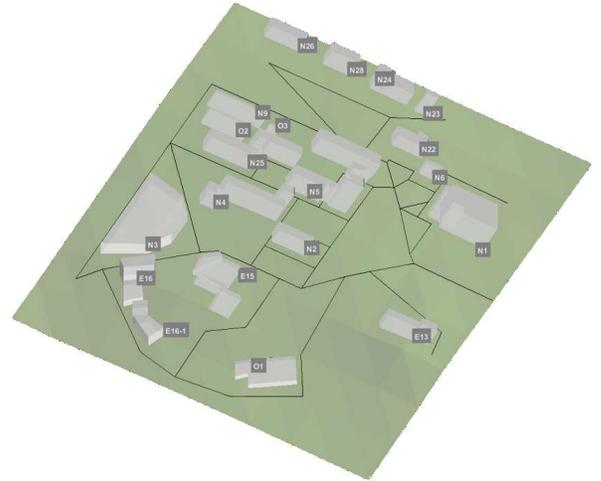}} \label{subfig:3-b}
		\end{array}$
	\end{center}
	\caption{3D map around Korea Advanced Institute of Science and Technology, South Korea.}
	\label{fig:3}
\end{figure}

\section{WiThRay Channel Simulator}
\label{sec3}
This section provides details on the WiThRay procedure. Section~\ref{sec3_1} explains a 3D map data that is imported for the RT algorithm. Section~\ref{sec3_2} covers the BE algorithm, the scattering criteria to adjust the scattering area, and the MIMO calibration method in WiThRay. In Section~\ref{sec3_3}, we define the geometrical parameters that are given by the BE algorithm. The CIR model and RIS response model used in WiThRay are given in Sections~\ref{sec3_4} and \ref{sec3_5}, respectively. Finally, Section~\ref{sec3_6} defines the output of WiThRay, which consists of the discrete sampled channel and the orthogonal frequency division multiplexing (OFDM) channel.

\begin{figure}
	\centering
	\begin{center}$
		\begin{array}{c}
			\subfloat[Simulation result with UTD calibration on refletion, transition, and shadow regions.]{\includegraphics[width=0.65\columnwidth]{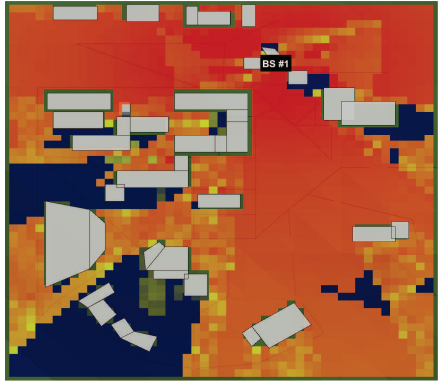} \label{subfig:4-a}} \\
			\subfloat[Simulation result with UTD calibration only on shadowed region.]{\includegraphics[width=0.65\columnwidth]{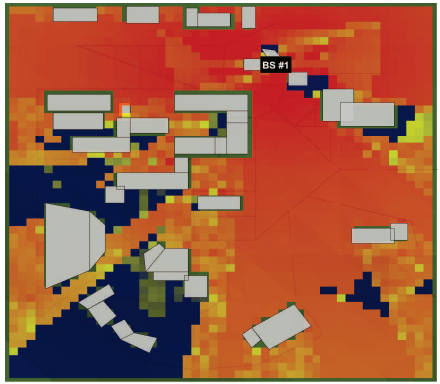} \label{subfig:4-b}}\\
			\subfloat{\includegraphics[width=0.8\columnwidth]{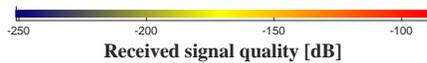} \label{subfig:4-c}}
		\end{array}$
	\end{center}
	\caption{Effect of diffraction in reflection and transition regions. Maximum numbers of reflections and diffractions are one in the simulation.}
	\label{fig:4}
\end{figure}

\subsection{3D Map Data}
\label{sec3_1}
The RT algorithm runs with the 3D map data. This paper conducts the concrete experiment with the map around the east gate of Korea Advanced Institute of Science and Technology (KAIST). The reference map is taken from the commercial website.\footnote{https://map.naver.com/} The 3D map includes various objects, e.g., buildings and vehicles. Buildings are static objects, and vehicles are objects that move or stop. The objects in the 3D map are the composite of hexahedrons. A hexahedron unit is not restricted to rectangular surfaces, increasing the flexibility of 3D modeling. By using more hexahedrons for 3D object modeling, the rendering accuracy can grow with the trade-off of increased complexity for the RT algorithm.

The terrain of map consists of consecutive triangular facets. Each point of a facet has the altitude information of the terrain. Roads are paved on the terrain, determining the routes for pedestrians and vehicles. A full sketch of the 3D map around KAIST is shown in Fig.~\ref{fig:3}. Considering the connectivity of roads, WiThRay automatically generates the routes for pedestrians and vehicles, where the temporal correlation of the channel engages with the spatial correlation. The discrete points on the route of a pedestrian are the consecutive locations of an mUE in the simulation. The mSOs have their own speeds that determine the moving distance per unit time, causing the Doppler shift.

\subsection{Considered Ray Tracing Methodology}
\label{sec3_2}
\subsubsection{Reflection and diffraction based on BE algorithm}
The RT methodology is dependent on the ray propagation scenario defined in the RT algorithm. When it comes to finding the reflection path, there is a commonly used approach. However, various approaches are used to determine the diffracting path.

Signals that diffract at the edge propagate through reflection, transition, and shadow regions defined in Fig.~\ref{fig:2}. Responses of diffracting signals that propagate to all directions can be evaluated using the GTD or UTD. However, as shown in Fig.~\ref{fig:4}, tracing all diffracting paths toward all regions is not markedly distinct from tracing only the diffracting paths toward the shadow region. Therefore, WiThRay's inverse algorithm, the BE algorithm, only traces the diffracting paths in the shadow region where only diffracting signals exist.

The diffracting path passing through the shadow region can be interpreted as a bypass path for a blocked signal. To account for this, the BE algorithm initially considers only reflections and explores all paths. It then checks whether bypassing paths exist for the objects that cause blockage and whether those paths exist on the Keller's cone. The process of the BE algorithm is summarized as follows:

\begin{enumerate}
	\item[i)] Find the coordinates of symmetrically flipped points of the Rx point for all existing surfaces and check whether the flipped Rx points are connected to the Tx point through reflecting surfaces.
	\item[ii)] Check whether the blockage occurs in the reflecting path and list the objects where the blockage occurs.
	\item[iii)] Explore whether there are diffracting points bypassing the identified blocking objects.
	\item[iv)] After finding the diffracting points, check again to see whether the blockage occurs even for the bypassing paths.
\end{enumerate}

\begin{figure}
	\centering
	\begin{center}$
		\begin{array}{c}
			\subfloat[Single reflection.]{\includegraphics[width=0.65\columnwidth]{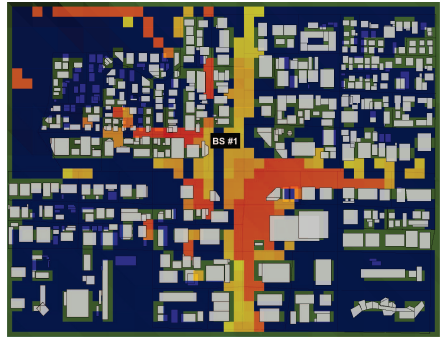} \label{subfig:5_a}}\\
			\subfloat[Single diffraction.]{\includegraphics[width=0.65\columnwidth]{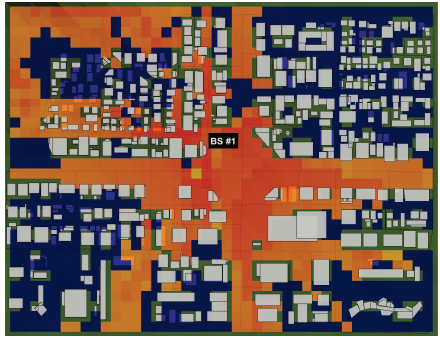} \label{subfig:5_b}} \\
			\subfloat[Single reflection and single diffraction.]{\includegraphics[width=0.65\columnwidth]{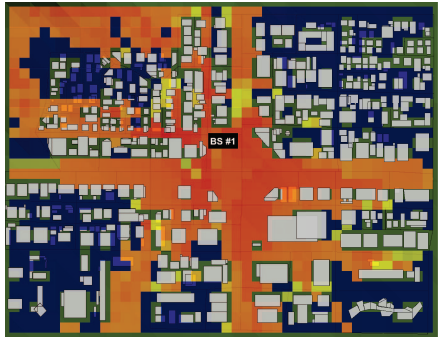} \label{subfig:5_c}}\\
			\subfloat[Double reflections and double diffractions.]{\includegraphics[width=0.65\columnwidth]{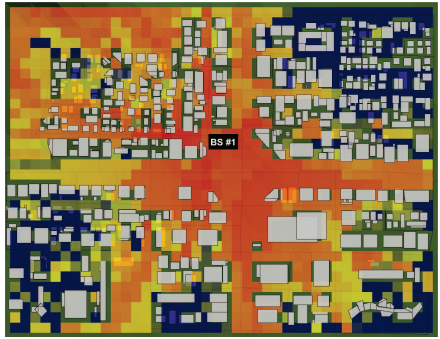} \label{subfig:5_d}}\\
			\subfloat{\includegraphics[width=0.8\columnwidth]{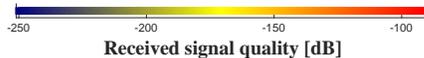} \label{subfig:5_e}}
		\end{array}$
	\end{center}
	
	\caption{Simulation results with different numbers of reflections and diffractions.}
	\label{fig:5}
\end{figure}

The BE algorithm is fully generalized such that WiThRay finds out all possible channel paths satisfying the Fermat's principle within the predefined numbers of reflection and diffraction while, theoretically, there is no limit on the number of reflecting and diffracting points in obtaining the channel path the RF signal travels. As shown in Fig.~\ref{fig:5}, the simulation accuracy improves by considering more reflection and diffraction paths. By increasing these paths, however, the path-searching complexity increases significantly. In the case of reflection, the number of paths to check is $M^L$ for $L$ reflections with $M$ surfaces. To enable realistic 3D environment designs with large amounts of objects, it is critical to reduce the number of cases to be checked before starting the path calibration.

\begin{figure}
	\centering
	\includegraphics[width=0.65\columnwidth]{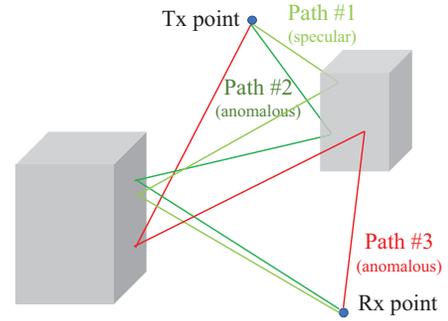}
	\caption{Specular and anomalous channel paths. Path \#2 is the anomalous path related with the specular Path \#1. Path \#2 has a meaningful effect on the channel when comparing with the anomalous Path \#3 that propagates to some random direction.}
	\label{fig:6}
\end{figure}

Some conditions on the reflection and diffraction help dramatically decrease the number of cases to check. The conditions used in WiThRay are summarized as follows:

\begin{enumerate}
	\item[\romannumeral1)] Any two consecutive reflecting surfaces must face each other since the line connecting them cannot go through the backside of any of them.
	\item[\romannumeral2)] The reflecting point must be within the reflecting surface.
	\item[\romannumeral3)] The diffracting point must be within the diffracting edge.
	\item[\romannumeral4)] The diffracting path is the bypassing path, which is created on the blocking object. Therefore, the diffraction must occur on the blocking objects.
	\item[\romannumeral5)] Most Tx and Rx antenna pairs at the same terminals are located close to each other, which makes them experience similar trajectories on common objects.
\end{enumerate}
These conditions are simple but powerful enough to decrease the simulation time effectively. The conditions {\romannumeral1})-{\romannumeral3}) coincide with the ray tube of VT method. However, the condition {\romannumeral4}) makes the BE algorithm different form the inverse algorithm with the VT method. Since the BE algorithm only finds the diffracting paths on blockage, the number of paths to be checked for the diffraction largely decreases, resulting in significant reduction in simulation time. In the MIMO scenario, the condition {\romannumeral5}) helps to remove redundant objects to be checked in the RT algorithm.

\subsubsection{Scattering}
\begin{figure}
	\centering
	\begin{center}$
		\begin{array}{c}
			\subfloat[Tx antenna at (-50,-10,40) and Rx antenna at (50,-10,1).]{\includegraphics[width=\columnwidth]{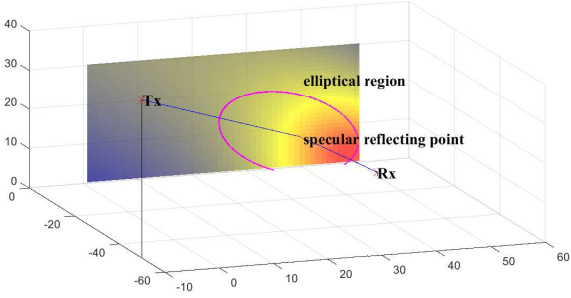} \label{subfig:7_a}}\\
			\subfloat[Tx antenna at (-40,-10,40) and Rx antenna at (50,-20,1).]{\includegraphics[width=\columnwidth]{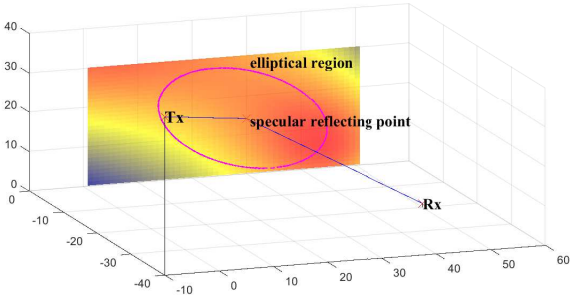} \label{subfig:7_b}}
		\end{array}$
	\end{center}
	\caption{Path gain of scattering path corresponding to each reflecting point. The elliptical region is an intersection of the surface and the 3D ellipsoid whose radius is defined as: 3 meters + distance between Tx and flipped Rx. (a) The dominant paths are in the elliptical region. (b) The ellipsoid does not contain all effective channel paths.}
	\label{fig:7}
\end{figure}

While anomalous paths can occur without any interaction with specular paths as in Fig.~\ref{fig:6}, these anomalous paths are usually much weaker compared to the specular paths and the anomalous paths related to the specular paths. To identify the anomalous paths, the PG-based method examines the grid points located in close proximity to the specular path. Each grid point represents the point where an anomalous path meets the surface. The length of the anomalous path is confined by the specular path. The scattering area, where the grid points are included, is the intersection of the surface and the 3D ellipsoid as in Fig.~\ref{fig:7}. The 3D ellipsoid has the Tx point and symmetrically flipped Rx point as its focal points \cite{Degli:2004,Zhang:2022}.

\begin{figure}
	\centering
	\begin{center}$
		\begin{array}{c}
			\subfloat[Focal point of specular reflecting paths.]{\includegraphics[width=0.75\columnwidth]{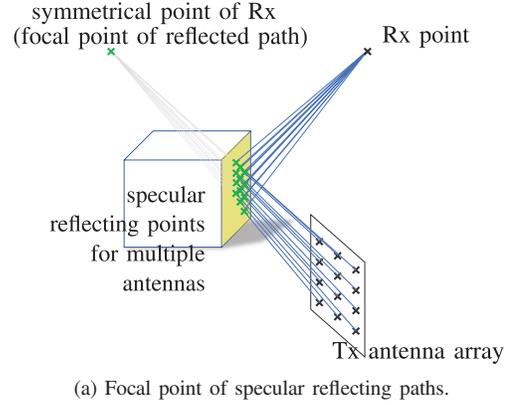} \label{subfig:8-a}}\\
			\subfloat[Focal point of anomalous scattering paths.]{\includegraphics[width=0.75\columnwidth]{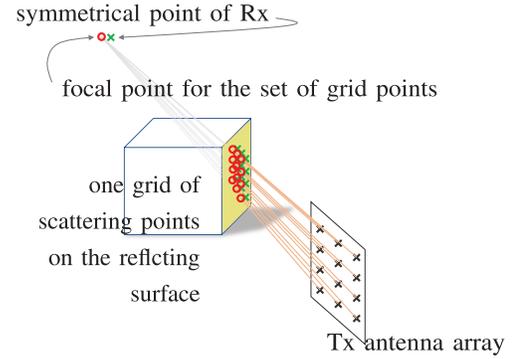} \label{subfig:8-b}}
		\end{array}$
	\end{center}
	\caption{Focal point mismatch of scattering points caused by the common grid interval. (a) Symmetrically flipped Rx point is the focal point of specular reflecting points (green x-dots) from multiple antennas, where specular paths are colored by blue. (b) Scattering grid points of anomalous paths shift the focal point.}
	\label{fig:8}
\end{figure}

To evaluate the effect of scattering, the path gain should be evaluated for all grid points in the ellipsoid. The path gain is obtained by the pathloss in trajectory, partial energy projected to the grid point, and EM response of reflection or diffraction. However, the channel paths bounded in the ellipsoid might not be the dominant channel paths. Fig.~\ref{fig:7} shows the two different cases of scattering. The dominant scattering paths belong to the elliptical region in Fig.~\ref{fig:7}~\subref{subfig:7_a} while the ellipsoid does not contain the dominant paths in Fig.~\ref{fig:7}~\subref{subfig:7_b}. Thus, WiThRay dynamically adjusts the radius of ellipsoid such that most of dominant paths belong to the elliptical region.

\begin{figure}
	\centering
	\begin{center}$
		\begin{array}{c}
			\subfloat[Anchor points obtained from the first Tx antenna.]{\includegraphics[width=0.75\columnwidth]{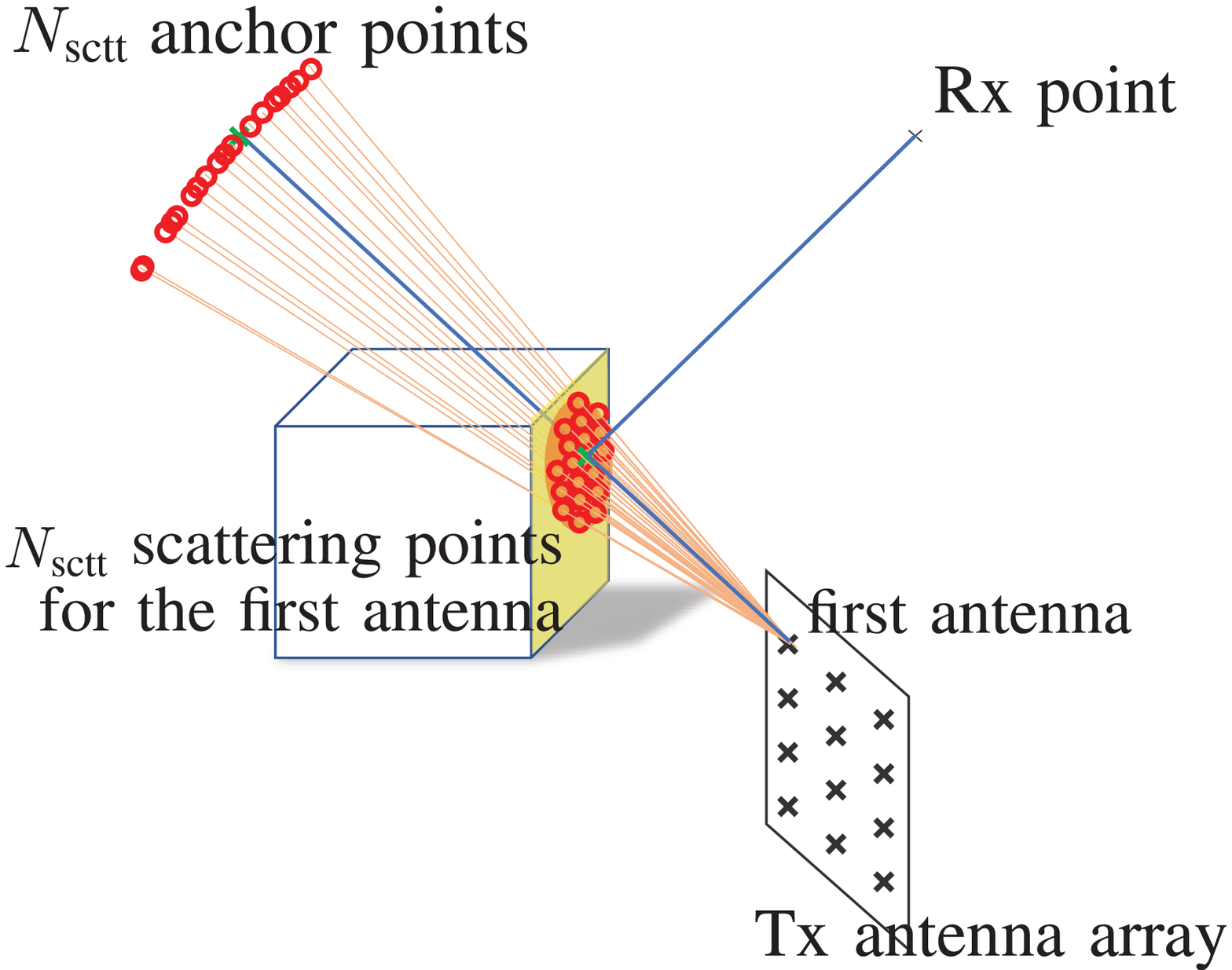} \label{subfig:9-a}}\\
			\subfloat[Scattering point for the second Tx antenna.]{\includegraphics[width=0.75\columnwidth]{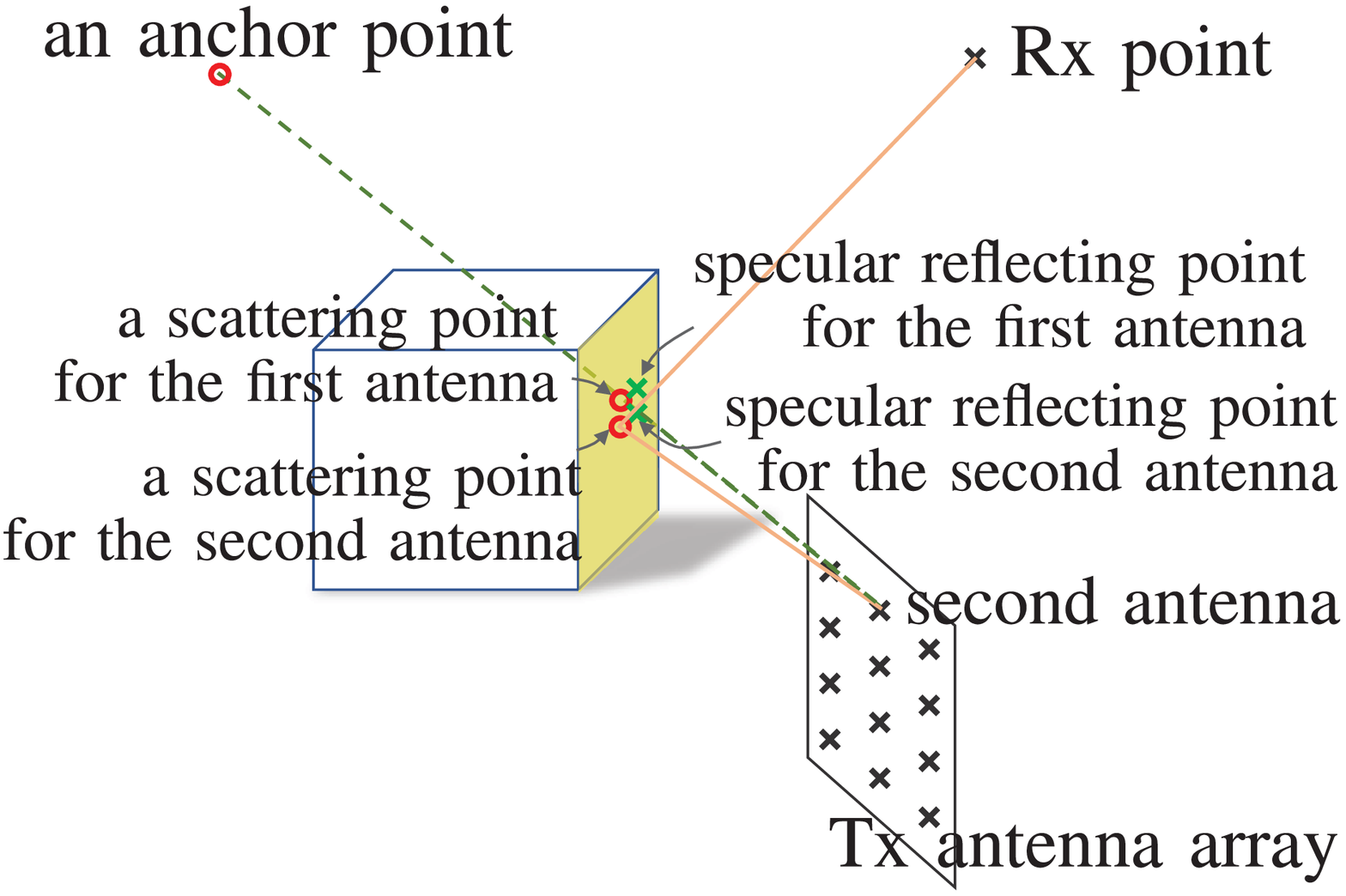} \label{subfig:9-b}}
		\end{array}$
	\end{center}
	\caption{Scattering points calibration using anchor points. (a) Anchor points obtained from the first Tx antenna are designed to have the same distance from the first antenna. (b) Scattering point interval for the second Tx antenna calibrated by the anchor point. The anchor point and second Tx antenna are connected by the green dotted line.}
	\label{fig:9}
\end{figure}

When it comes to MIMO systems, designing the scattering grid interval should be more delicate. Each antenna has its own reflecting point, as in Fig.~\ref{fig:8}~\subref{subfig:8-a}, which means each reflecting point has its own grid for the scattering. When the reflecting points from different antennas have the common grid interval for the scattering, the focal point of the scattering path becomes shifted, as shown in Fig.~\ref{fig:8}~\subref{subfig:8-b}. Because of this mismatch between the symmetrically flipped Rx point and the focal point, the RF signals of multiple antennas lose their pattern and do not act as a planar wave. Since the grid is just a concept to implement the scattering effect within the RT algorithm, and the actual RF signal, even in the scattering path, moves as a planar wave if the distance between the Tx and Rx points is sufficiently long, the scattering grid interval in Fig.~\ref{fig:8}~\subref{subfig:8-b} should be redesigned.

\begin{figure}
	\centering
	\begin{center}$
		\begin{array}{c}
			\subfloat[Experiment scenario.]{\includegraphics[width=0.9\columnwidth]{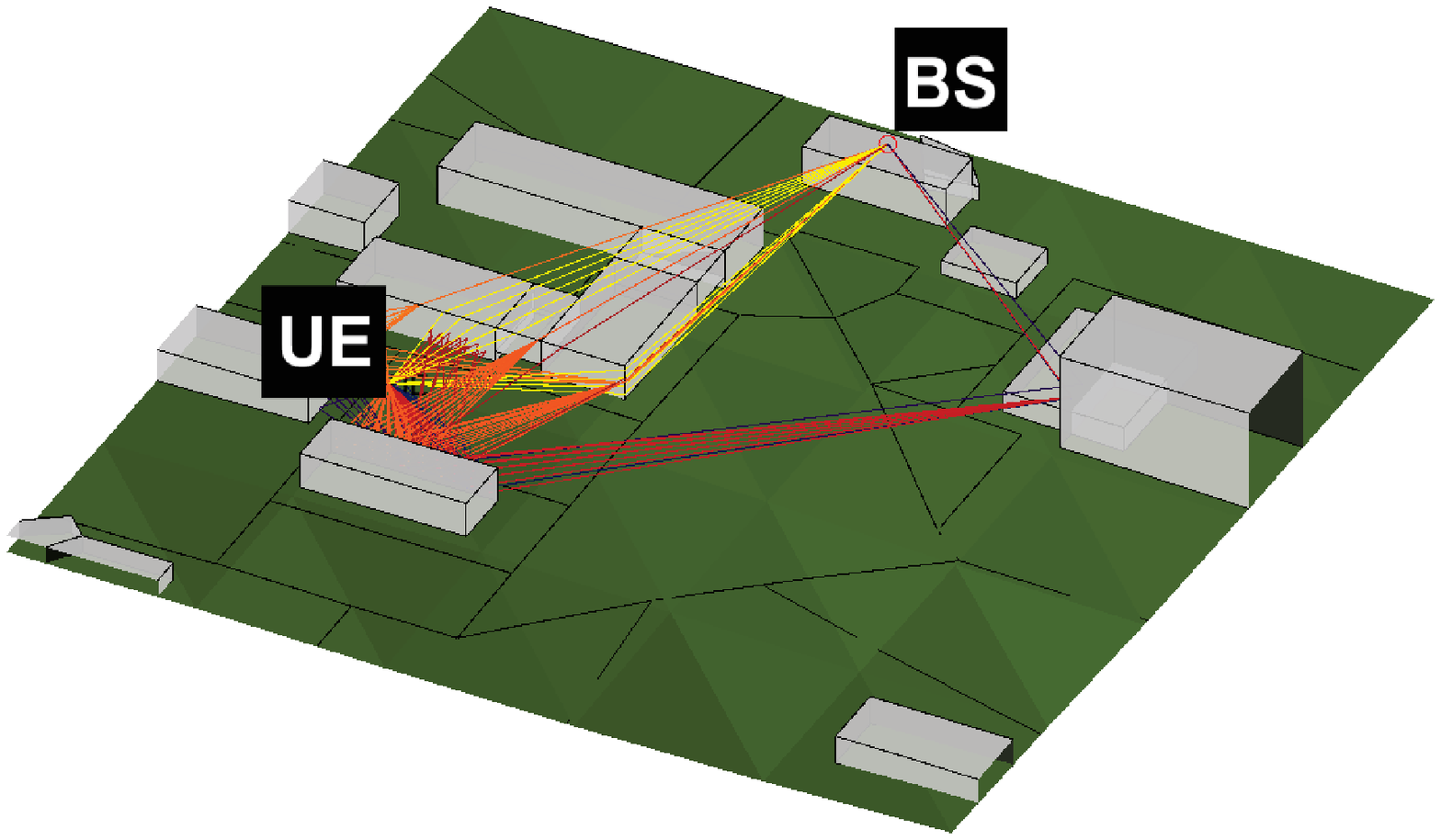} \label{subfig:10-a}}\\
		\end{array}$
	\end{center}
	\begin{center}$
		\begin{array}{cc}
			\subfloat[UPA beam pattern without the MIMO calibration.]{\includegraphics[width=0.47\columnwidth]{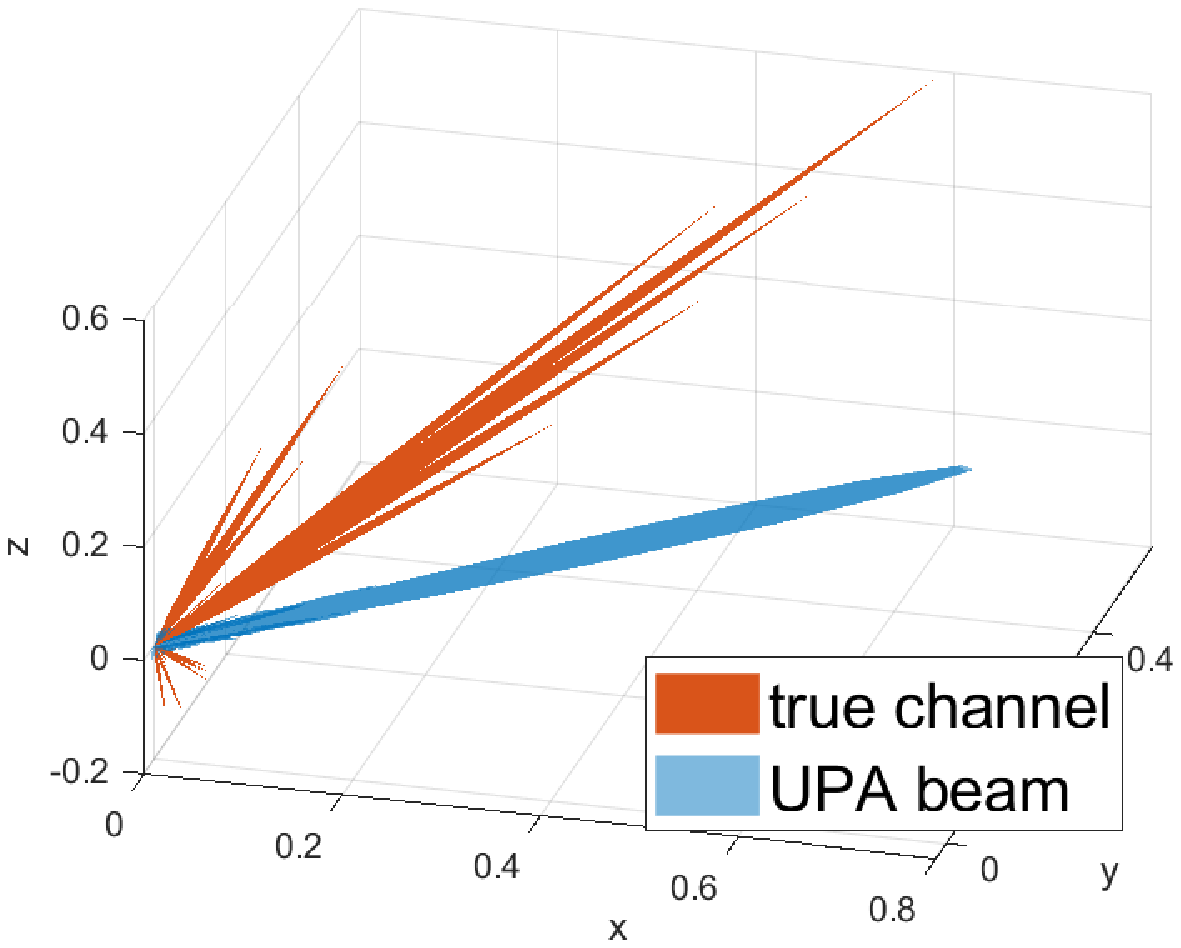} \label{subfig:10-b}}&
			\subfloat[UPA beam pattern with the MIMO calibration.]{\includegraphics[width=0.47\columnwidth]{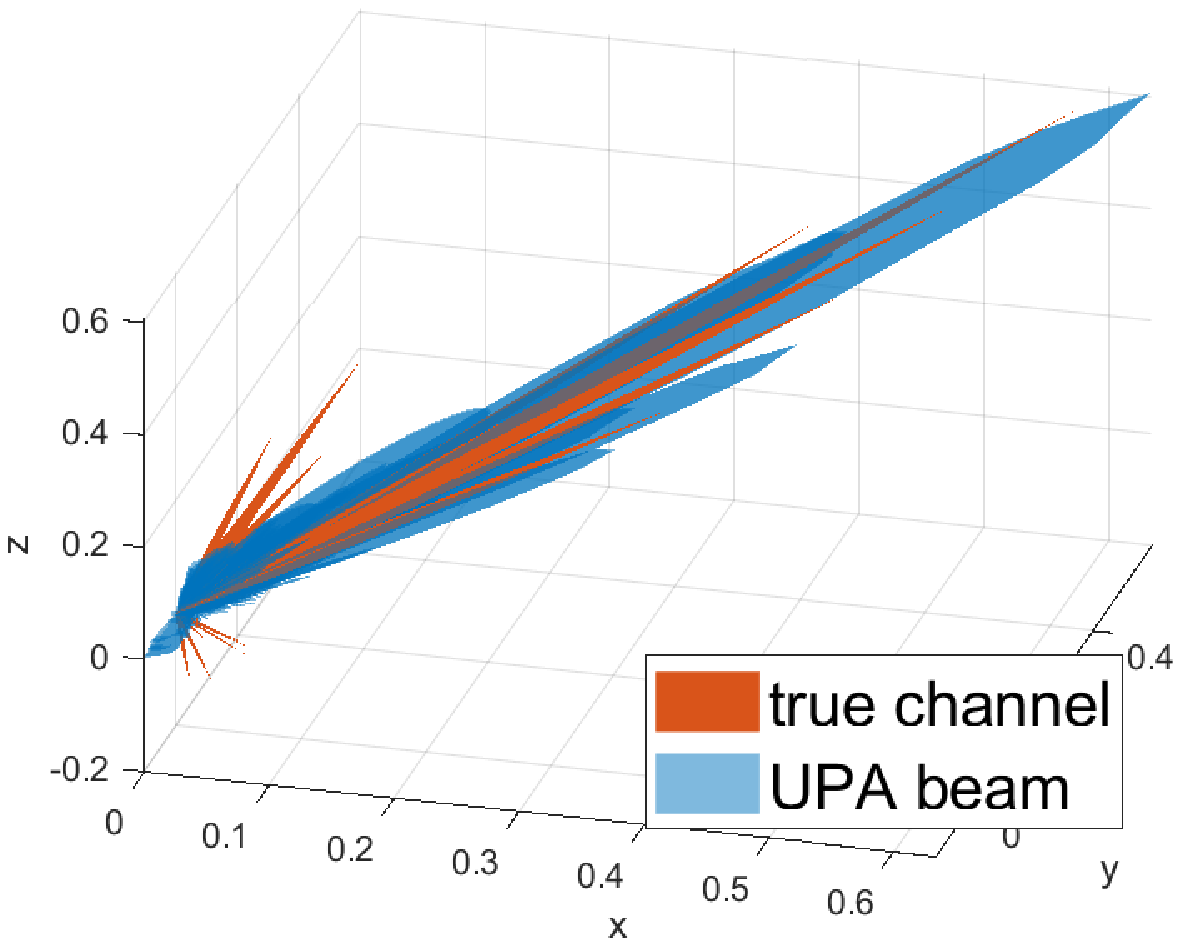} \label{subfig:10-c}}
		\end{array}$
	\end{center}
	\caption{The MIMO calibration in WiThRay. (a) The BS and the UE are connected with a large number of scattering paths. (b) True channel shows the dominant channel paths that actually come in. However, the beam pattern using UPA at the UE does not indicate the direction of true channel. (c) The MIMO calibration adjust the UPA beam pattern such that it aligns to the direction of true channel.}
	\label{fig:10}
\end{figure}

To manipulate the planar wave for the scattering paths, WiThRay judiciously calibrates the grid points for the Tx (or Rx) antennas. First, we introduce the anchor points in Fig.~\ref{fig:9} to calibrate the grid points. The scattering grid points for the first Tx antenna are used to set the anchor points, where each anchor point determines the grid points for other Tx antennas. The anchor point is the point on the line connecting the first Tx antenna point and the scattering grid point for the first Tx antenna. There are $N_\text{sctt}$ anchor points for $N_\text{sctt}$ scattering grid points, where the distances from the first antenna to the anchor points are all set to be equal to the length of the specular path as in Fig.~\ref{fig:9}~\subref{subfig:9-a}, which is the reference for $N_\text{sctt}$ scattering paths. This is because there should be no difference in the propagation loss for the specular path and the scattering paths.

The $n_\text{sctt}$-th anchor point is to evaluate the length of the $n_\text{sctt}$-th scattering path that is defined as the sum of distances between 1) the Rx antenna point and the $n_\text{sctt}$-th scattering grid point and 2) the $n_\text{sctt}$-th scattering grid point and the Tx antenna point, which is indicated by the red line in Fig.~\ref{fig:9}~\subref{subfig:9-b} considering the second Tx antenna as an example. When the $n_\text{sctt}$-th scattering path length is equal to the distance between the $n_\text{sctt}$-th anchor point and the second Tx antenna point, i.e., the green dotted line in Fig.~\ref{fig:9}~\subref{subfig:9-b}, the $n_\text{sctt}$-th scattering path acts as a planer wave, since the $n_\text{sctt}$-th anchor point makes the traveling distances of RF signals from Tx antennas uniformly different. As shown in Fig.~\ref{fig:10}, the beam pattern constructed at the user equipment (UE) using uniform planar array (UPA) could align with the direction of incoming signals only when the MIMO calibration manipulates the scattering paths.

The scattering points are shifted from the specular reflecting and diffracting points. Thus, the scattering points can be interpreted as additional reflecting and diffracting points in anomalous trajectories. All the specular or anomalous reflecting and diffracting points are saved sequentially as important points (IPs) in WiThRay. Each IP has its coordinate, response type, i.e., reflection or diffraction, and object identification.

\subsection{Evaluation of Geometrical Parameters}
\label{sec3_3}

The lines connecting reflecting and diffracting points from the Tx point to the Rx point are the trajectories of channel paths. After the RT algorithm calibrates all the trajectories of channel paths, WiThRay evaluates the directions of the propagation and polarization. The directions of the propagation and polarization are denoted by the $3\times 1$ directional unit vectors $\bd_{p,q}$ and $\boldsymbol{\varphi}_{p,q}$, whose elements are the directions in the axes of the 3D Cartesian coordinate system. The directional unit vectors $\bd_{p,q}$ and $\boldsymbol{\varphi}_{p,q}$ are identified with the indices of the trajectory $p\in\{1,\dots,P\}$ and the IP $q\in\{1,\dots,Q_p\}$, where $P$ is the number of channel path trajectories, and $Q_p$ is the number of IPs in the $p$-th trajectory. Specifically, $\boldsymbol{\varphi}_{p,q}^{(E)}$ and $\boldsymbol{\varphi}_{p,q}^{(H)}$ are the directions of electric and magnetic fields. Since the IPs in the $p$-th trajectory vary with different positions of Tx and Rx antennas, $\bd_{p,q}$, $\boldsymbol{\varphi}_{p,q}^{(E)}$, and $\boldsymbol{\varphi}_{p,q}^{(H)}$ depend on the Tx and Rx antennas. Thus, the directions of propagation and polarization are denoted as $\bd_{p,q}^{\mathfrak{n}_t, \mathfrak{n}_r}$, $\boldsymbol{\varphi}_{p,q}^{(E),\mathfrak{n}_t, \mathfrak{n}_r}$, and $\boldsymbol{\varphi}_{p,q}^{(H),\mathfrak{n}_t, \mathfrak{n}_r}$. The indices of the $(n_\text{hor}^t,n_\text{ver}^t)$-th Tx and the $(n_\text{hor}^r,n_\text{ver}^r)$-th Rx antennas are $\mathfrak{n}_t=(n_\text{hor}^t-1)N_\text{ver}^t+n_\text{ver}^t$ and $\mathfrak{n}_r=(n_\text{hor}^r-1)N_\text{ver}^r+n_\text{ver}^r$, where the size of Tx antenna array is $N_\text{hor}^t\times N_\text{ver}^t$ and the size of Rx antenna array is $N_\text{hor}^r\times N_\text{ver}^r$. The $q$-th propagating distance $r_{p,q}^{\mathfrak{n}_t, \mathfrak{n}_r}$ between the $q$-th and $(q+1)$-th IPs in the $p$-th trajectory also depends on the positions of the Tx and Rx antennas. The overall length of the $p$-th trajectory between the $\mathfrak{n}_t$-th Tx antenna and $\mathfrak{n}_r$-th Rx antenna is $r_p^{\mathfrak{n}_t, \mathfrak{n}_r}=\sum_{q=1}^{Q_p}r_{p,q}^{\mathfrak{n}_t, \mathfrak{n}_r}$. Then, the propagation delay $\tau_p$ of the $p$-th trajectory can be evaluated by the traveling distance as $\tau_p^{\mathfrak{n}_t, \mathfrak{n}_r}=r_p^{\mathfrak{n}_t, \mathfrak{n}_r}/c$, where $c$ is the speed of light.

\begin{figure}
	\centering
	\includegraphics[width=0.9\columnwidth]{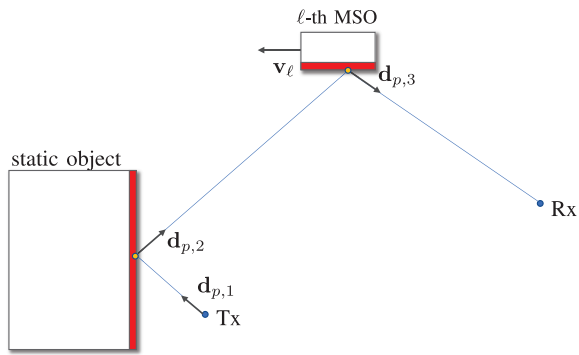}
	\caption{The trajectory with the points reflecting on static and mobile objects is depicted. The $\ell$-th mSO moves with $\bv_\ell$. The $q$-th propagation of the $p$-th trajectory has the propagation direction $\bd_{p,q}$. The IP index of the $\ell$-th mSO is $q(\ell)=2$ in this example.}
	\label{fig:11}
\end{figure}

The direction of propagation $\bd_{p,q}^{\mathfrak{n}_t,\mathfrak{n}_r}$ is related to the angles of azimuth/zenith of departure/arrival (AoD/ZoD/AoA/ZoA), where the AoD/ZoD/AoA/ZoA on the $q$-th IP of $p$-th trajectory are denoted, respectively, as $\theta_{p,q}^t$, $\phi_{p,q}^t$, $\theta_{p,q}^r$, and $\phi_{p,q}^r$. For the geometrical channel model, the path gain $\alpha_p$ and the array steering vector $\ba(\theta,\phi)$ define the response of a single channel path as $\bH_p(t)=\alpha_p\ba(\theta_{p,Q_p}^r,\phi_{p,Q_p}^r)\ba^\mathrm{H}(\theta_{p,1}^t,\phi_{p,1}^t)$. The array steering vector of the $N_\text{hor}^t\times N_\text{ver}^t$ uniform planar array (UPA) antenna at the Tx is determined by $\theta_{p,q}^t$ and $\phi_{p,q}^t$ as
\begin{align}\label{eq:SteerVec}
	\ba(\theta_{p,q}^t,\phi_{p,q}^t)&=\begin{bmatrix}1 & e^{j\Delta\psi_\text{hor}^t} & \cdots & e^{j(N_\text{hor}^t-1)\Delta\psi_\text{hor}^t}\end{bmatrix}^\mathrm{T} \otimes \nonumber \\
	&\begin{bmatrix}1 & e^{j\Delta\psi_\text{ver}^t} & \cdots & e^{j(N_\text{ver}^t-1)\Delta\psi_\text{ver}^t}\end{bmatrix}^\mathrm{T},
\end{align}
with $\Delta\psi_\text{hor}^t=2\pi d^t\cos(\theta_{p,q}^t)\sin(\phi_{p,q}^t)/\lambda$ and $\Delta\psi_\text{ver}^t=2\pi d^t\cos(\phi_{p,q}^t)/\lambda$, where $d^t$ is the interval of two adjacent antennas, and $\lambda$ is the wavelength of the carrier frequency. Different from the geometrical model that exploits the array response vector in \eqref{eq:SteerVec}, the RT-based simulators including WiThRay do not rely on the array steering vector for the channel modeling. WiThRay directly exploits the channel phases for all Tx and Rx antenna pairs to build the CIR, which makes it suitable for near-field experiments. As will become clear in Section~\ref{sec3_4}, the resulting channel still exhibits a steering pattern similar to the array response vector as evaluated in Section~\ref{sec4_3}.

Another important geometrical parameter of the channel is the Doppler frequency. Each $(\mathfrak{n}_t,\mathfrak{n}_r)$-th antenna pair has different trajectories of the channel paths, which makes the Doppler frequency of the $(\mathfrak{n}_t,\mathfrak{n}_r)$-th antenna pair unique. To simplify the notations, we neglect the indices $\mathfrak{n}_t$ and $\mathfrak{n}_r$ in the following definition. The Doppler shift in the $p$-th trajectory defined in WiThRay is given as\cite{Bilibashi:2020}
\begin{align}\label{eq:doppler}
	\bar{f}_p=f_c\left(1-\frac{1}{c}\left(\bar{v}_p+\sum_{\ell\in\cL_p^r\cup\cL_p^d}\bar{v}_{\ell,p}\right)\right),
\end{align}
where $f_c$ is the carrier frequency, $\bar{v}_p$ and $\bar{v}_{\ell,p}$ are the relative speeds of the mUE and the $\ell$-th mSO. The relative speed of the mUE in the $p$-th trajectory is $\bar{v}_p=\bv\cdot\bd_{p,Q_p}$, where $\bv$ is the velocity of the mUE, and $\bd_{p,Q_p}$ is the propagating direction of the $p$-th trajectory of the UE. The velocity $\bv$ of the mUE is a $3\times 1$ directional vector whose elements are the speeds of mUE in the 3D Cartesian coordinate.

The relative speed of the $\ell$-th mSO in the $p$-th trajectory is obtained as $\bar{v}_{p,\ell}=\bv_\ell\cdot(\bd_{p,q(\ell)}-\bd_{p,q(\ell)+1})$ where $\bv_\ell$ denotes the velocity of the $\ell$-th mSO. The index $q(\ell)$ refers to the IP index at the $\ell$-th mSO, where the directions of propagation $\bd_{p,q(\ell)}$ and $\bd_{p,q(\ell)+1}$ are the incoming and outgoing directions related to the $\ell$-th mSO, as shown in Fig.~\ref{fig:11}. The total sum of relative speeds, $\bar{v}_p+\sum_{\ell\in\cL_p^r\cup\cL_p^d}\bar{v}_{\ell,p}$, is the change of length of the $p$-th trajectory per unit time, where $\cL_p^r$ and $\cL_p^d$ are the sets of mSOs under the reflection and diffraction. If there is no mSO, \eqref{eq:doppler} can be simplified as $\bar{f}_p=f_c(1-\bar{v}_p/c)$, the well known form of the Doppler frequency.

\subsection{Channel Impulse Response Modeling}
\label{sec3_4}
In WiThRay, the channel from the $\mathfrak{n}_t$-th Tx antenna to the $\mathfrak{n}_r$-th Rx antenna is determined by their positions and polarization directions. The CIR from the $\mathfrak{n}_t$-th Tx antenna to the $\mathfrak{n}_r$-th Rx antenna is, therefore, constructed as follows:
\begin{align}\label{eq:ChannelModel}
	\left[\bH(t,\mathfrak{t})\right]&_{\mathfrak{n}_t, \mathfrak{n}_r}= \sum_{p=1}^P\xi^t(\bd_{p,1}^{\mathfrak{n}_t, \mathfrak{n}_r}(\mathfrak{t}))\xi^r(\bd_{p,Q_p}^{\mathfrak{n}_t, \mathfrak{n}_r}(\mathfrak{t})) \nonumber \\
	&\times\exp\left(-j2\pi \bar{f}_p^{\mathfrak{n}_t, \mathfrak{n}_r}(\mathfrak{t})\tau_p^{\mathfrak{n}_t, \mathfrak{n}_r}(\mathfrak{t})\right)\delta(t-\tau_p^{\mathfrak{n}_t, \mathfrak{n}_r}(\mathfrak{t})) \nonumber \\
	&\times\prod_{q=1}^{Q_p+1}\frac{\sqrt{P_0/4\pi}}{\eta(r_{p,q}^{\mathfrak{n}_t, \mathfrak{n}_r}(\mathfrak{t}),\bar{f}_{p}^{\mathfrak{n}_t, \mathfrak{n}_r}(\mathfrak{t}))} \nonumber \\
	&\times\prod_{q\in\cQ_r}\zeta^r(\bd_{p,q}^{\mathfrak{n}_t, \mathfrak{n}_r}(\mathfrak{t}),\bd_{p,q+1}^{\mathfrak{n}_t, \mathfrak{n}_r}(\mathfrak{t}),\boldsymbol{\varphi}_{p,q}^{(E),\mathfrak{n}_t, \mathfrak{n}_r}(\mathfrak{t})) \nonumber \\
	&\times\prod_{q\in\cQ_d}\zeta^d(\bd_{p,q}^{\mathfrak{n}_t, \mathfrak{n}_r}(\mathfrak{t}),\bd_{p,q+1}^{\mathfrak{n}_t, \mathfrak{n}_r}(\mathfrak{t}),\boldsymbol{\varphi}_{p,q}^{(E),\mathfrak{n}_t, \mathfrak{n}_r}(\mathfrak{t})),
\end{align}
which is the channel at time $t$ for the RF signal starting at time $\mathfrak{t}$. The CIR in \eqref{eq:ChannelModel} follows the geometrical optics \cite[Ch. 13]{balanis1999advanced}, where the antenna radiation pattern $\xi(\cdot)$, reflecting response $\zeta^r(\cdot)$, and diffracting response $\zeta^d(\cdot)$ are independent and coherently combined in a product from \eqref{eq:ChannelModel}. The directions of propagation $\bd_{p,q}^{\mathfrak{n}_t,\mathfrak{n}_r}(\mathfrak{t})$ and polarization $\boldsymbol{\varphi}_{p,q}^{(E),\mathfrak{n}_t,\mathfrak{n}_r}(\mathfrak{t})$, the traveling distance $r_{p,q}^{\mathfrak{n}_t,\mathfrak{n}_r}(\mathfrak{t})$, the delay $\tau_{p,q}^{\mathfrak{n}_t,\mathfrak{n}_r}(\mathfrak{t})$, and the Doppler frequency $\bar{f}_p^{\mathfrak{n}_t,\mathfrak{n}_r}(\mathfrak{t})$ of the $p$-th trajectory are functions of the starting time $\mathfrak{t}$. The pathloss $\eta(r,f_c)$ is the function of the atmospheric absorption, which depends on the carrier frequency\cite{series2017guidelines}. In the following descriptions, we simplify the notations by omitting the common indices $\mathfrak{n}_t$, $\mathfrak{n}_r$, and $\mathfrak{t}$.

The functions $\xi^t(\cdot)$ and $\xi^r(\cdot)$ are the radiation patterns at the Tx and Rx antennas determined by the actual antenna design. The dipole antenna is a popular antenna structure, where its radiation pattern is modeled as
\begin{align}
	\xi_\text{dipole}(\bd_{p,q})=\lVert\bd_{p,q}\times\boldsymbol{\varphi}_\text{dipole}\rVert,
\end{align}
where $\boldsymbol{\varphi}_\text{dipole}$ represents the direction of dipole antenna. There are various types of antenna structure with more complicated radiation patterns. They are, however, still a function of the propagation $\bd_{p,q}$. Thus, WiThRay is compatible with all types of antenna structure by properly fixing the antenna direction $\boldsymbol{\varphi}$.

The reflecting response function $\zeta^r(\cdot)$ is the EM response determined by the directions of propagation and polarization.  Because the propagating direction $\bd_{p,q}$ coming into the $q$-th IP and the direction $\bd_{p,q+1}$ going out to the $(q+1)$-th IP are both related to the reflection on the $q$-th IP, the reflecting response function at the $q$-th IP can be modeled as
\begin{align}\label{eq:ReflectiveResponse}
	\zeta^r(\bd_{p,q},\bd_{p,q+1},\boldsymbol{\varphi}_{p,q}^{(E)})=&E_{r,s}(\bd_{p,q},\bd_{p,q+1}) \nonumber \\ &\times E_{r,p}(\bd_{p,q},\bd_{p,q+1},\boldsymbol{\varphi}_{p,q}^{(E)}),
\end{align}
where $E_{r,s}(\cdot)$ is the scattering response and $E_{r,p}(\cdot)$ is the loss caused by the polarization.
The directive scattering model is widely used for the scattering response \cite{Fuschini:2015,Wang:2016,WirelessInSite,Tian:2016,Pascual:2016}. If the $p$-th trajectory is an anomalous scattering path, the $p$-th trajectory passes a grid point around the $p^\ast$-th specular trajectory. Then, the directive scattering model is given as \cite{Degli:2007}
\begin{align}\label{eq:DirectiveModel}
	E_{r,s}^2(\bd_{p,q},\bd_{p,q+1})=&\left(E_s^\text{max}(\bd_{p,q})\right)^2 \nonumber \\ 
	&\times\left(\frac{1+\cos\varPsi_{p,q+1|p^\ast}}{2}\right)^{\alpha_R},
\end{align}
where the alignment angle $\varPsi_{p,q+1|p^\ast}$ is defined as $\varPsi_{p,q+1|p^\ast}=\cos^{-1}(\bd_{p,q+1}\cdot\bd_{p^\ast,q+1})$. The direction $\bd_{p^\ast,q+1}$ is the $(q+1)$-th propagating direction on the $p^\ast$-th specular trajectory. The exponent $\alpha_R\in\mathbb{N}$ in \eqref{eq:DirectiveModel} decides the width of the scattering lobe. The maximum amplitude of the scattering lobe in \eqref{eq:DirectiveModel} is calculated as
\begin{align}\label{eq:MaxScatt}
	(E_s^\text{max}(\bd_{p,q}))^2=\frac{dS\cos\vartheta_{p,q}}{F_{\alpha_R}(\vartheta_{p,q})},
\end{align}
with the incident angle $\vartheta_{p,q}=\cos^{-1}(\bd_{p,q}\cdot\bn_{p,q}^\text{surf})$. Note that $\bn_{p,q}^\text{surf}$ is the normal vector of the surface where the $q$-th IP on the $p$-th trajectory is placed. The scattering effect of a single grid area is normalized with the grid size $dS$. The scaling factor $F_{\alpha_R}(\vartheta_{p,q})$ in \eqref{eq:MaxScatt} is given as 
\begin{align}
	F_{\alpha_R}(\vartheta_{p,q})&=\frac{1}{2^{\alpha_R}}\sum_{j=0}^{\alpha_R}\begin{pmatrix}\alpha_R \\ j\end{pmatrix}\frac{2\pi}{j+1}\times \nonumber \\
	&\left[\cos(\vartheta_{p,q})\sum_{w=0}^\frac{j-1}{2}\begin{pmatrix}2w \\ w\end{pmatrix}\frac{\sin^{2w}\vartheta_{p,q}}{2^{2w}}\right]^{\frac{1-(-1)^j}{2}}.
\end{align}

To derive the polarization loss $E_{r,p}(\cdot)$, first note that the induced current on the reflecting surface of an object is determined by the projection of the incident magnetic field onto the surface. The current induced on the surface causes the vector potential field, where the polarized RF signal radiates. The reflecting power of the polarized wave decreases as the inner product of the two directional unit vectors, i.e., the induced current direction $\boldsymbol{\varphi}_{p,q}^{(J)}$ and the reflecting direction $\bd_{p,q+1}$, increases. Considering these two effects, the polarization loss in the reflecting response is expressed as follows\cite[Ch. 7]{balanis1999advanced}
\begin{align}
	E_{r,p}^2(\bd_{p,q}&,\bd_{p,q+1},\boldsymbol{\varphi}_{p,q}^{(E)})=\nonumber\\
	&\chi_{r,p}\frac{\lVert\boldsymbol{\varphi}_{p,q}^{(H)}\times\bn_{p,q}^\text{surf}\rVert \lVert\boldsymbol{\varphi}_{p,q}^{(J)}\times\bd_{p,q+1}\rVert+\kappa_r}{1+\kappa_r},
\end{align}
where the direction of the incident magnetic field is $\boldsymbol{\varphi}_{p,q}^{(H)}=\bd_{p,q}\times\boldsymbol{\varphi}_{p,q}^{(E)}$, and the direction of the induced current on the reflecting surface is $\boldsymbol{\varphi}_{p,q}^{(J)}=(\bn_{p,q}^\text{surf}\times\boldsymbol{\varphi}_{p,q}^{(H)})/\lVert\bn_{p,q}^\text{surf}\times\boldsymbol{\varphi}_{p,q}^{(H)}\rVert$. The reflecting coefficient $\chi_{r,p}$ is the EM coefficient identifying the material property of the $p$-th IP, and the polarization coefficient $\kappa_r$ determines the degree of polarization effect on the total reflecting response.

\begin{figure}
	\centering
	\begin{center}$
		\begin{array}{c}
			\subfloat[3D view.]{\includegraphics[width=0.75\columnwidth]{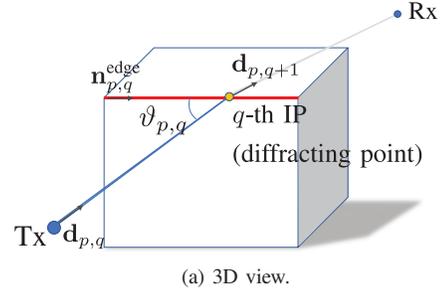} \label{subfig:12-a}}\\
			\subfloat[Side view.]{\includegraphics[width=0.75\columnwidth]{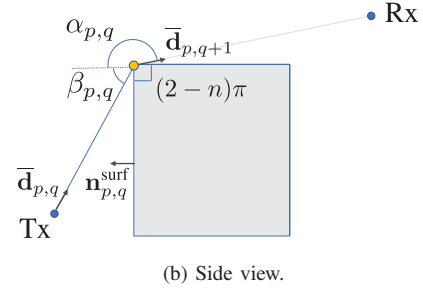} \label{subfig:12-b}}
		\end{array}$
	\end{center}
	\caption{The diffracting point on the edge. (a) The incident angle of the incident wave and the diffracting edge is $\vartheta_{p,q}$, which is defined on the 3D coordinate. (b) There are parameters defined on the 2D coordinate; $\alpha_{p,q}$, $\beta_{p,q}$, and $n$. The 2D angles of diffracting and incident waves are denoted as $\alpha_{p,q}$ and $\beta_{p,q}$, respectively. The wedge of object has angle $(2-n)\pi$ with $n=1.5$, since the object has a rectangular shape.}
	\label{fig:12}
\end{figure}

The diffracting response $\zeta^d(\cdot)$ is also a function of the propagating directions $\bd_{p,q}$, $\bd_{p,q+1}$, and the polarizing direction $\boldsymbol{\varphi}_{p,q}^{(E)}$. The diffracting response function can be mathematically expressed as \cite{Kouyoumijan:1974}
\begin{align}\label{eq:DiffractiveResponse}
	\zeta^d(\bd_{p,q},&\bd_{p,q+1},\boldsymbol{\varphi}_{p,q}^{(E)}) \nonumber \\
	&=\left(\lvert E_d^{(s)}(\bd_{p,q},\bd_{p,q+1})\rvert^2\cdot\lvert\boldsymbol{\varphi}_{p,q}^{(s)}\cdot\boldsymbol{\varphi}_{p,q+1}^{(s)}\rvert^2\right. \nonumber \\
	&\ + \left.\lvert E_d^{(h)}(\bd_{p,q},\bd_{p,q+1})\rvert^2\cdot\lvert\boldsymbol{\varphi}_{p,q}^{(h)}\cdot\boldsymbol{\varphi}_{p,q+1}^{(h)}\rvert^2\right)^{1/2},
\end{align}
where $E_d^{(s)}$ and $E_d^{(h)}$ are the diffracting effects for the soft and hard polarizations, respectively. The hard polarization $\boldsymbol{\varphi}_{p,q}^{(h)}$ is a polarization direction perpendicular to the diffracting edge, which is obtained as $\boldsymbol{\varphi}_{p,q}^{(h)}=(\bn_{p,q}^\text{edge}\times\bd_{p,q})/\lVert\bn_{p,q}^\text{edge}\times\bd_{p,q}\rVert$. The soft polarization can be evaluated as $\boldsymbol{\varphi}_{p,q}^{(s)}=(\bd_{p,q}\times\boldsymbol{\varphi}_{p,q}^{(h)})/\lVert\bd_{p,q}\times\boldsymbol{\varphi}_{p,q}^{(h)}\rVert$.

The polarization direction of the incident electric field can be decomposed as $\boldsymbol{\varphi}_{p,q}^{(E)}=(\boldsymbol{\varphi}_{p,q}^{(E)}\cdot\boldsymbol{\varphi}_{p,q}^{(s)})\cdot\boldsymbol{\varphi}_{p,q}^{(s)}+(\boldsymbol{\varphi}_{p,q}^{(E)}\cdot\boldsymbol{\varphi}_{p,q}^{(h)})\cdot\boldsymbol{\varphi}_{p,q}^{(h)}$, and the hard and soft polarization components of the incident electric field experience individual diffractions in the UTD. The UTD formulates the soft and hard diffractions as \cite{Kouyoumijan:1974}
\begin{align}\label{eq:UTD} 
	E_{d}^{(s,h)}&(\bd_{p,q},\bd_{p,q+1})=\frac{-\sqrt{\lambda}e^{-j\frac{\pi}{4}}}{2\pi \sin\vartheta_{p,q}} \nonumber \\ &\times\left[\cot\left(\frac{\pi+(\alpha_{p,q}-\beta_{p,q})}{2n}\right)\right. \nonumber \\
	&\qquad\qquad\times F\left[\frac{2\pi r_{p,q}}{\lambda}a^+(\alpha_{p,q}-\beta_{p,q})\sin\vartheta_{p,q}\right] \nonumber\\ &+\cot\left(\frac{\pi-(\alpha_{p,q}-\beta_{p,q})}{2n}\right) \nonumber \\
	&\qquad\qquad\times F\left[\frac{2\pi r_{p,q}}{\lambda}a^-(\alpha_{p,q}-\beta_{p,q})\sin\vartheta_{p,q}\right] \nonumber \\
	&\mp\left\{\cot\left(\frac{\pi+(\alpha_{p,q}+\beta_{p,q})}{2n}\right)\right. \nonumber \\
	&\qquad\qquad\times F\left[\frac{2\pi r_{p,q}}{\lambda}a^+(\alpha_{p,q}+\beta_{p,q})\sin\vartheta_{p,q}\right] \nonumber \\
	&\qquad+\cot\left(\frac{\pi-(\alpha_{p,q}+\beta_{p,q})}{2n}\right) \nonumber \\
	&\qquad\qquad\times \left.\left.F\left[\frac{2\pi r_{p,q}}{\lambda}a^-(\alpha_{p,q}+\beta_{p,q})\sin\vartheta_{p,q}\right]\right\} \right],
\end{align}
where the auxiliary functions $F(x)$ and $a^\pm(\beta)$ in \eqref{eq:UTD} are
\begin{align}
	F(x) = 2je^{jx}\sqrt{x}\int_{\sqrt{x}}^{\infty}e^{-jw}dw
\end{align}
and
\begin{align}
	a^\pm(\beta)=2\cos^2\left(\frac{2n\pi N^\pm-\beta}{2}\right).
\end{align}
The integers $N^\pm$ are $N^+=[\![(\pi+\beta)/2\pi n]\!]$ and $N^-=[\![(-\pi+\beta)/2\pi n]\!]$, where $[\![\cdot]\!]$ means a round integer. The $q$-th angles of the $p$-th trajectory; $\vartheta_{p,q}$, $\alpha_{p,q}$, and $\beta_{p,q}$, are described in Fig.~\ref{fig:12}. The incident angle is $\vartheta_{p,q}=\cos^{-1}(\bd_{p,q}\cdot\bn^\text{edge}_{p,q})$. The diffracting and incident angles in the 2D coordinate are $\alpha_{p,q}=\cos^{-1}(\overline{\bd}_{p,q+1}\cdot\bn_{p,q}^\text{surf})$ and $\beta_{p,q}=\cos^{-1}(\overline{\bd}_{p,q}\cdot\bn_{p,q}^\text{surf})$, respectively, where $\overline{\bd}_{p,q}$ and $\overline{\bd}_{p,q+1}$ are the projected vectors of $\bd_{p,q}$ and $\bd_{p,q+1}$, respectively, on the surface normal to the edge $\bn_{p,q}^\text{edge}$. The wedge number $n$ in Fig.~\ref{fig:12}~\subref{subfig:12-b} is 1.5, which means the object has a rectangular shape.

To derive the polarization direction $\boldsymbol{\varphi}_{p,q+1}^{(E)}$ of the diffracting RF signal, we can express \eqref{eq:DiffractiveResponse} as
\begin{align}
	\zeta^d(\bd_{p,q},\bd_{p,q+1},\boldsymbol{\varphi}_{p,q}^{(E)})=\left(\lvert D^{(s)}\rvert^2+\lvert D^{(h)}\rvert^2\right)^{1/2},
\end{align}
where $D^{(s)}= E_d^{(s)}(\bd_{p,q},\bd_{p,q+1})\cdot(\boldsymbol{\varphi}_{p,q}^{(s)}\cdot\boldsymbol{\varphi}_{p,q+1}^{(s)})$ and $D^{(h)}= E_d^{(h)}(\bd_{p,q},\bd_{p,q+1})\cdot(\boldsymbol{\varphi}_{p,q}^{(h)}\cdot\boldsymbol{\varphi}_{p,q+1}^{(h)})$. The diffracting polarization direction $\boldsymbol{\varphi}_{p,q+1}^{(E)}$ is defined as $\boldsymbol{\varphi}_{p,q+1}^{(E)}=(D^{(s)}\boldsymbol{\varphi}_{p,q+1}^{(s)}+D^{(h)}\boldsymbol{\varphi}_{p,q+1}^{(h)})/\sqrt{\lvert D^{(s)}\rvert^2+\lvert D^{(h)}\rvert^2}$.

By defining the path gain $\alpha_p^{\mathfrak{n}_t,\mathfrak{n}_r}(\mathfrak{t})$ as the combined effect of the first and the last three lines in \eqref{eq:ChannelModel}, the CIR can be compactly written as
\begin{align}\label{eq:SimpleModel}
	&\left[\bH(t,\mathfrak{t})\right]_{\mathfrak{n}_t,\mathfrak{n}_r}=\sum_{p=1}^P\alpha_p^{\mathfrak{n}_t,\mathfrak{n}_r}(\mathfrak{t})\nonumber \\
	&\qquad\times \exp\left(-j2\pi\bar{f}_p^{\mathfrak{n}_t,\mathfrak{n}_r}(\mathfrak{t})\tau_p^{\mathfrak{n}_t,\mathfrak{n}_r}(\mathfrak{t})\right)\delta(t-\tau_p^{\mathfrak{n}_t,\mathfrak{n}_r}(\mathfrak{t})),
\end{align}
where the path gain $\alpha_p^{\mathfrak{n}_t,\mathfrak{n}_r}(\mathfrak{t})$ is
\begin{align}\label{eq:PathGain}
	\alpha_p^{\mathfrak{n}_t,\mathfrak{n}_r}&(\mathfrak{t})= \xi^t(\bd_{p,1}^{\mathfrak{n}_t, \mathfrak{n}_r}(\mathfrak{t}))\xi^r(\bd_{p,Q_p}^{\mathfrak{n}_t, \mathfrak{n}_r}(\mathfrak{t})) \nonumber \\
	&\times\prod_{q=1}^{Q_p+1}\frac{\sqrt{P_0/4\pi}}{\eta(r_{p,q}^{\mathfrak{n}_t, \mathfrak{n}_r}(\mathfrak{t}),\bar{f}_{p}^{\mathfrak{n}_t, \mathfrak{n}_r}(\mathfrak{t}))} \nonumber \\
	&\times\prod_{q\in\cQ_r}\zeta^r(\bd_{p,q}^{\mathfrak{n}_t, \mathfrak{n}_r}(\mathfrak{t}),\bd_{p,q+1}^{\mathfrak{n}_t, \mathfrak{n}_r}(\mathfrak{t}),\boldsymbol{\varphi}_{p,q}^{(E),\mathfrak{n}_t, \mathfrak{n}_r}(\mathfrak{t})) \nonumber \\
	&\times\prod_{q\in\cQ_d}\zeta^d(\bd_{p,q}^{\mathfrak{n}_t, \mathfrak{n}_r}(\mathfrak{t}),\bd_{p,q+1}^{\mathfrak{n}_t, \mathfrak{n}_r}(\mathfrak{t}),\boldsymbol{\varphi}_{p,q}^{(E),\mathfrak{n}_t, \mathfrak{n}_r}(\mathfrak{t})).
\end{align}
In the following sections, the simplified CIR in \eqref{eq:SimpleModel} is used instead of the full CIR in \eqref{eq:ChannelModel}.

\textit{\textbf{Remark:}} The reflecting response in \eqref{eq:ReflectiveResponse} and diffracting response in \eqref{eq:DiffractiveResponse} do not consider the EM properties of materials. However, the reflective and diffractive coefficients, e.g., permittivity and permeability, depend on the EM material properties, which may affect the phase shift during the reflection and diffraction in practice. The design purpose of WiThRay is to describe the temporal and spatial variation patterns in the RF channel, which do not depend much on the accurate phase responses in the reflection and diffraction. The patterns are generated by the relative delay of the multiple channel paths. The phase of the channel model in \eqref{eq:ChannelModel}, therefore, only deals with the periodicity embedded in the effective channel in \eqref{eq:SimpleModel}.

\begin{figure}
	\centering
	\includegraphics[width=1\columnwidth]{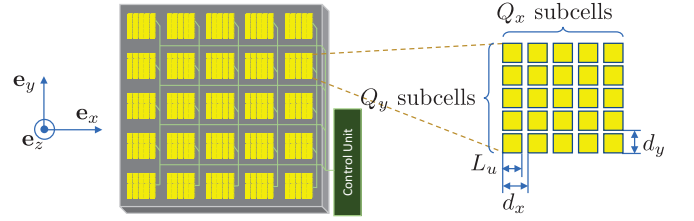}
	\caption{Discrete-cell type RIS.}
	\label{fig:13}
\end{figure}

\subsection{RIS Response Model}
\label{sec3_5}
The RIS implemented in WiThRay is a discrete-cell type where each panel comprises a number of subcells as in Fig.~\ref{fig:13}. To steer the reflective wave on the electrical panel, the phase variation on the panel must change continuously as the signal impinges upon different points on the panel's surface. The discrete-cell type RIS is the implementation solution for the phase variation on the surface. The magnitude response 	$\zeta_{\Vert,\text{RIS}}(\cdot)$ and the phase response $\zeta_{\angle,\text{RIS}}(\cdot)$ on the discrete-cell type RIS are given as \cite{Najafi:2021}
\begin{align}\label{eq:RISresponse}
	&\zeta_{\Vert,\text{RIS}}(\bd_{p,q},\bd_{p,q+1},\boldsymbol{\varphi}_{p,q}^{(E)})=\frac{\sqrt{4\pi}L_u^2}{\lambda}c_{\boldsymbol{\varphi}}\nonumber \\
	&\times\left\lVert\begin{bmatrix}[\tilde{\bd}_{p,q}]_1\sqrt{1-[\tilde{\bd}_{p,q}]_3^2}\cos{\varphi_m}-[\tilde{\bd}_{p,q}]_1[\tilde{\bd}_{p,q}]_3\sin{\varphi_m} \nonumber \\
		\sqrt{1-[\tilde{\bd}_{p,q}]_3^2}\sin{\varphi_m}+[\tilde{\bd}_{p,q}]_3\cos{\varphi_m}\end{bmatrix}\right\rVert\nonumber \\
	&\times\text{sinc}\left(\frac{\pi L_u A_x}{\lambda}\right)
	\text{sinc}\left(\frac{\pi L_u A_y}{\lambda}\right)\nonumber \\
	&\times\frac{\sin\left(\frac{\pi Q_x d_x}{\lambda}(A_x-A_x^\ast)\right)}{\sin\left(\frac{\pi d_x}{\lambda}(A_x-A_x^\ast)\right)}
	\frac{\sin\left(\frac{\pi Q_x d_x}{\lambda}(A_y-A_y^\ast)\right)}{\sin\left(\frac{\pi d_x}{\lambda}(A_y-A_y^\ast)\right)},
\end{align}
\begin{align}
	&\zeta_{\angle,\text{RIS}}(\bd_{p,q},\bd_{p,q+1},\boldsymbol{\varphi}_{p,q}^{(E)})=\frac{\pi}{2}+\frac{\pi d_x(A_x-A_x^\ast)}{\lambda} \nonumber \\
	&\qquad\qquad\qquad\qquad\qquad\qquad\qquad+\frac{\pi d_y(A_y-A_y^\ast)}{\lambda},
\end{align}
where the projection coefficient $c_{\varphi}$ in \eqref{eq:RISresponse} is modeled as
\begin{align}
	c_{\boldsymbol{\varphi}}=\frac{A_z}{\sqrt{A_{xy}^2+A_z^2}},
\end{align}
with $A_{xy}=[\tilde{\bd}_{p,q}]_1\cos{\varphi_m}+[\tilde{\bd}_{p,q}]_2\sin\varphi_m$ and $A_z=[\tilde{\bd}_{p,q}]_3$. The relative directional vector $\tilde{\bd}_{p,q}$ is a rotated vector that is obtained as $\tilde{\bd}_{p,q}=[\mathbf{e}_x\ \mathbf{e}_y\ \mathbf{e}_z]^\mathrm{T}\bd_{p,q}$ with the RIS coordinate basis vectors $\mathbf{e}_x$, $\mathbf{e}_y$, and $\mathbf{e}_z$ described in Fig.~\ref{fig:13}. The intervals between subcells in x-axis $d_x$ and in y-axis $d_y$, the number of subcells in x-axis $Q_x$ and in y-axis $Q_y$, and the length of subcell $L_u$ are also given in Fig.~\ref{fig:13}. The directional changes in x-axis and y-axis are $A_x=[\tilde{\bd}_{p,q}]_1+[\tilde{\bd}_{p,q+1}]_1$ and $A_y=[\tilde{\bd}_{p,q}]_2+[\tilde{\bd}_{p,q+1}]_2$, and $A_x^\ast$ and $A_y^\ast$ are intended directions of propagation, which determine the directivity of scattering. The angle of projected magnetic field is $\varphi_m=\tan^{-1}([\boldsymbol{\varphi}_{p,q}^{(H)}]_2/[\boldsymbol{\varphi}_{p,q}^{(H)}]_1)$. The total RIS response can be represented as
\begin{align}
	\zeta_\text{RIS}&(\bd_{p,q},\bd_{p,q+1},\boldsymbol{\varphi}_{p,q}^{(E)})=\nonumber \\
	&\zeta_{\Vert,\text{RIS}}(\bd_{p,q},\bd_{p,q+1},\boldsymbol{\varphi}_{p,q}^{(E)})e^{{j\zeta_{\angle,\text{RIS}}(\bd_{p,q},\bd_{p,q+1},\boldsymbol{\varphi}_{p,q}^{(E)})}}.
\end{align}

To combine the RIS response in the CIR, WiThRay should first obtain the path gain, delay, and Doppler shift between the BS and RIS, as well as between the RIS and UE. The RIS channel from the $\mathfrak{n}_t$-th Tx antenna to the $\mathfrak{n}_r$-th Rx antenna through the $\mathfrak{n}_i$-th RIS cell can be obtained as
\begin{align}\label{eq:CIR_ris}
	[\boldsymbol{\cH}(t,\mathfrak{t})&]_{\mathfrak{n}_t,\mathfrak{n}_i,\mathfrak{n}_r}=\sum_{p=1}^P\sum_{r=1}^R\alpha_p^{\mathfrak{n}_t,\mathfrak{n}_i}(\mathfrak{t})\beta_r^{\mathfrak{n}_i,\mathfrak{n}_r}(\tau_p^{\mathfrak{n}_t,\mathfrak{n}_i}(\mathfrak{t})) \nonumber \\
	&\times\zeta_\text{RIS}(\bd_{p,Q_p+1},\bd_{r,1},\boldsymbol{\varphi}_{p,Q_p+1}^{(E)}) \nonumber \\
	&\times\exp\left[-j2\pi\left\{\bar{f}_p^{\mathfrak{n}_t,\mathfrak{n}_i}(\mathfrak{t})+\bar{f}_r^{\mathfrak{n}_i,\mathfrak{n}_r}(\tau_p^{\mathfrak{n}_t,\mathfrak{n}_i}(\mathfrak{t}))\right\}\right. \nonumber \\
	&\qquad\qquad\qquad\times\left.\{\tau_p^{\mathfrak{n}_t,\mathfrak{n}_i}(\mathfrak{t})+\tau_r^{\mathfrak{n}_i,\mathfrak{n}_r}(\tau_p^{\mathfrak{n}_t,\mathfrak{n}_i}(\mathfrak{t}))\}\right] \nonumber \\
	&\times\delta\left\{\mathfrak{t}-\tau_p^{\mathfrak{n}_t,\mathfrak{n}_i}(\mathfrak{t})-\tau_r^{\mathfrak{n}_i,\mathfrak{n}_r}(\tau_p^{\mathfrak{n}_t,\mathfrak{n}_i}(\mathfrak{t}))\right\},
\end{align}
where the $p$-th path gain $\alpha_p^{\mathfrak{n}_t,\mathfrak{n}_i}$ between the BS and RIS and the $r$-th path gain $\beta_r^{\mathfrak{n}_i,\mathfrak{n}_r}$ between the RIS and UE can be obtained as in \eqref{eq:PathGain}. There are $P$ paths between the BS and RIS and $R$ paths between the RIS and UE. With \eqref{eq:CIR_ris}, the RIS-aided channel $\boldsymbol{\cH}$ can be evaluated in the same way as the direct channel $\bH$ in Section~\ref{sec3_4}.

\subsection{Discrete-Time Channel Model}
\label{sec3_6}
The CIR in Section~\ref{sec3_4} is the impulse response on the continuous-time domain. To obtain the discrete-time channel model, the pulse shaping filter should be considered. On the discrete-time domain, the transmit pulse shaping filter makes the digital signal transmittable, while mitigating the inter-symbol interference. The transmitted signal after the pulse shaping filter $p_\text{Tx}(t)$ is given as
\begin{align}
	x(t) = \sum_n x[n]p_\text{Tx}(t-nT_s),	
\end{align}
where $x[n]$ is the modulated Tx symbols, and $T_s$ is the sampling period. Assuming a noiseless case for simplicity, the received signal that undergoes the CIR in \eqref{eq:ChannelModel} is obtained as
\begin{align}\label{eq:RxSig}
	y(t)=\int x(\tau)h(t,\tau) d\tau.
\end{align}
In \eqref{eq:RxSig}, the simplified notation of the channel, $h(t,\tau)$, is $[\bH(t,\tau)]_{\mathfrak{n}_t,\mathfrak{n}_r }$ in \eqref{eq:SimpleModel}. The discrete sampled signal at time $t=mT_s$ after the receive pulse shaping filter $p_\text{Rx}(t)$ is given as
\begin{align}
	y[m]&=\left. y(t)\ast p_\text{Rx}(t+\tau_\text{sync})\right|_{t=mT_s}\nonumber \\
	&=\int y(u)p_\text{Rx}(mT_s-u+\tau_\text{sync})du\nonumber \\
	&=\iint\sum_n x[n] p_\text{Tx}(\tau-nT_s)h(u,\tau)\nonumber \\
	&\qquad\qquad\qquad\times p_\text{Rx}(mT_s-u+\tau_\text{sync})d\tau du\nonumber \\
	&=\sum_n x[n]\iint h(u,\tau)p_\text{Tx}(\tau-nT_s)\nonumber \\
	&\qquad\qquad\qquad\times p_\text{Rx}(mT_s-u+\tau_\text{sync})d\tau du.
\end{align}
Then, the time-variant (TV) discrete-time domain channel $\mathfrak{h}_\text{TV}[m,n]$ for the transmitted symbol $x[n]$ can be represented as
\begin{align}\label{eq:DiscreteChModel}
	\mathfrak{h}_\text{TV}&[m,n]=\iint h(u,\tau)p_\text{Tx}(\tau-nT_s)\nonumber \\
	&\qquad\qquad\quad\times p_\text{Rx}(mT_s-u+\tau_\text{sync})d\tau du\nonumber \\
	=&\int \sum_p\alpha_p(\tau)\exp(-j2\pi\bar{f}_p(\tau)\tau_p(\tau))\nonumber\\
	&\qquad\quad\times p_\text{Tx}(\tau-nT_s)p_\text{Rx}(mT_s-\tau_p(\tau)+\tau_\text{sync})d\tau\nonumber\\
	\stackrel{(\alpha)}{\approx}&\sum_p\alpha_p(nT_s)\exp(-j2\pi\bar{f}_p(nT_s)\tau_p(nT_s))\nonumber\\
	&\qquad\quad\times p((m-n)T_s-\tau_p(nT_s)+\tau_\text{sync}),
\end{align}
where the effective pulse shaping filter is $p(t)=p_\text{Tx}(t)\ast p_\text{Rx}(t)$, and ($\alpha$) is due to the assumption that $p_\text{Tx}(t)\approx 0$ for $t\notin[0,T_s)$, and $\alpha_p(t+nT_s)\approx \alpha_p(nT_s)$, $\bar{f}_p(t+nT_s)\approx\bar{f}_p(nT_s)$, and $\tau_p(t+nT_s)\approx \tau_p(nT_s)$ for $t\in[0,T_s)$.\footnote{For 2 nsec sampling period, i.e., 50 MHz sampling frequency, i) the maximum difference in the position of UE moving at 10 m/s is 0.2 $\mu$m, which means the maximum variation in the delay is limited to 0.667 fsec, ii) for 10m/s$^2$ acceleration with 50 MHz sampling frequency, the maximum variation in speed is 2 nm/s, and iii) when the carrier frequency is 2 GHz, the maximum Doppler shift would be 13.3 nHz for one sampling period. Since these values are relatively very small, we can regard the path gain, delay, and Doppler shift are constant.} The last three assumptions state that the channel gain $\alpha_p(\mathfrak{t})$, Doppler frequency $\bar{f}_p(\mathfrak{t})$, and delay $\tau_p(\mathfrak{t})$ do not change much during the sampling period.

When the CIR is time-invariant (TI), the discrete channel model in \eqref{eq:DiscreteChModel} becomes the convolution of the CIR and the pulse shaping filter as follows:
\begin{align}\label{eq:LTI}
	\mathfrak{h}_\text{TI}[m,n]&=\left. h(-t)\ast p_\text{Tx}(t-nT_s)\ast p_\text{Rx}(t+\tau_\text{sync})\right|_{t=mT_s}\nonumber \\
	&=\left.h(-t)\ast p(t-nT_s+\tau_\text{sync})\right|_{t=mT_s}\nonumber \\
	&=\sum_p \alpha_p\exp(-j2\pi\bar{f}_p\tau_p)p((m-n)T_s-\tau_p+\tau_\text{sync})\nonumber \\
	&=\mathfrak{h}_\text{TI}[m-n],
\end{align}
where the path gain $\alpha_p$, carrier frequency $\bar{f}_p$, and delay $\tau_p$ becomes independent of time $\mathfrak{t}$. The simplified notation $\mathfrak{h}[m,n]$ in the following paragraphs denotes the TV channel $\mathfrak{h}_\text{TV}[m,n]$ in \eqref{eq:DiscreteChModel}.

\begin{figure}
	\centering
	\begin{center}$
		\begin{array}{c}
			\subfloat[2.3 GHz carrier frequency.]{\includegraphics[width=0.7\columnwidth]{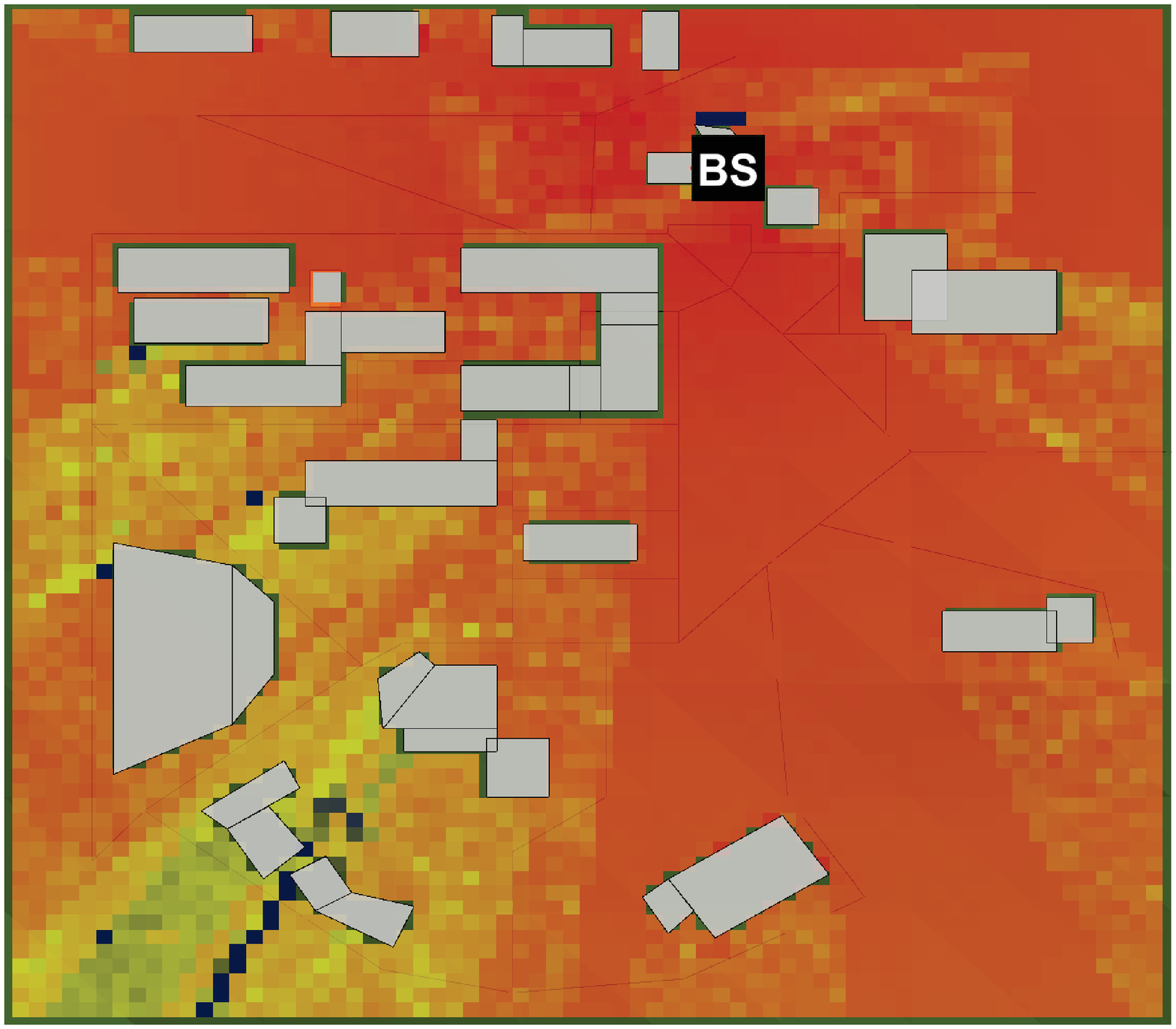} \label{subfig:14_a}}\\
			\subfloat[24 GHz carrier frequency.]{\includegraphics[width=0.7\columnwidth]{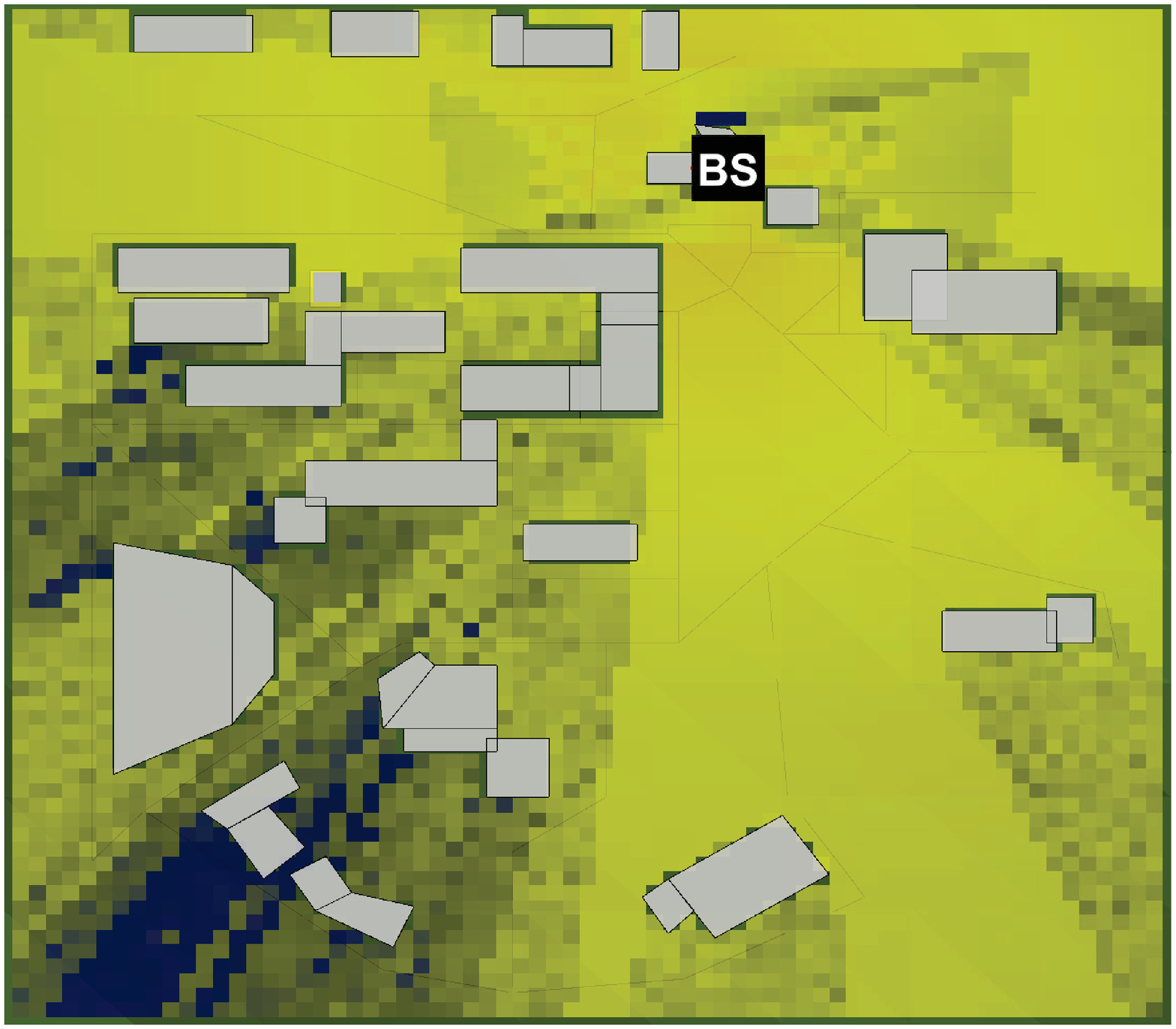} \label{subfig:14_b}} \\
			\subfloat{\includegraphics[width=0.9\columnwidth]{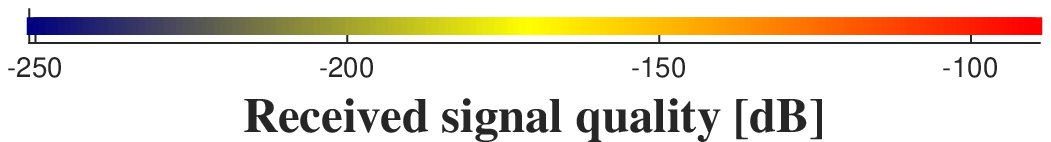} \label{subfig:14_c}}
		\end{array}$
	\end{center}
	
	\caption{Grid map of received signal quality. Note that the color scales are different in the two figures.}
	\label{fig:14}
\end{figure}

For the transmitting time index $n$, WiThRay evaluates the channel parameters $\alpha_p(nT_s)$, $\bar{f}_p(nT_s)$, and $\tau_p(nT_s)$ with given positions of Tx and Rx at $\mathfrak{t}=nT_s$. Among the CIRs starting at $\mathfrak{t}=nT_s$, the CIRs whose delay $\tau_p(nT_s)$ satisfies $\lceil\tau_p(nT_s)/T_s\rceil=m$ contribute to the TV discrete channel $\mathfrak{h}[m,n]$. The multi-tap channel model using \eqref{eq:DiscreteChModel} is given as
\begin{align}\label{eq:multitap}
	\bH=\begin{bmatrix}\mathfrak{h}[0,0] & 0 & 0 & \cdots \\
		\mathfrak{h}[1,0] & \mathfrak{h}[1,1] & 0      & \cdots \\
		\mathfrak{h}[2,0] & \mathfrak{h}[2,1] & \mathfrak{h}[2,2] & \cdots \\
		\vdots & \vdots & \vdots & \ddots \end{bmatrix}.
\end{align}
If there is no CIR satisfying $\lceil\tau_p(n'T_s)/T_s\rceil=m'$, the $(m',n')$-th tap channel $h[m',n']$ is zero. 

The OFDM channel with $K$ subcarriers can be obtained using the multi-tap channel model in \eqref{eq:multitap} with the cyclic prefix. After adding and removing the cyclic prefix with the length $N_c$, the $K\times K$ channel matrix $\tilde{\bH}=\bH'+\bH''$ is obtained, where the first summand $\bH'$ is a part of the multi-tap channel $\bH$ in \eqref{eq:multitap}, i.e., $\bH'=[\bH]_{N_c+1:N_c+K,N_c+1:N_c+K}$, and the second summand $\bH''$ is a part of the channel $\bH$ concatenated with the zero matrix, i.e., $\bH''= \left[\b0_{K,K-N_c},\ [\bH]_{N_c+1:N_c+K,1:N_c}\right]$. By conducting the inverse discrete Fourier transformation (IDFT) at the Tx and the discrete Fourier transformation (DFT) at the Rx, the OFDM channel can be written as
\begin{align}\label{eq:OFDMchannel}
	\boldsymbol{\cH}=\bF\tilde{\bH}\bF^\mathrm{H},
\end{align}
where $\bF$ is the $K\times K$ DFT matrix. The OFDM channel matrix in \eqref{eq:OFDMchannel} becomes a diagonal matrix only when $\tilde{\bH}$ is circulant. Since the multi-tap channel in \eqref{eq:multitap} is TV, the channel $\tilde{\bH}$ is not circulant in general, and there exists inter-carrier interference while the amount of interference varies with the number of significant channel taps in \eqref{eq:multitap}.

\begin{figure}
	\centering
	\includegraphics[width=0.7\columnwidth]{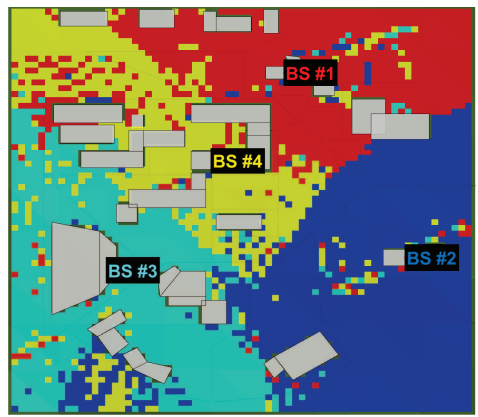}
	\caption{Service areas of 4 BSs are visualized by colored cubes. (BS \#1: red, BS \#2: blue, BS \#3: cyan, BS \#4: yellow)}
	\label{fig:15}
\end{figure}

\section{WiThRay's Features}
\label{sec4}
The proposed wireless channel simulator can provide various observations. The received signal quality, which is the ratio of the received signal power to transmit signal power when both the Tx and Rx are equipped with a single dipole antenna, is given on the grid points in Figs.~\ref{fig:14}~{(a)} and {(b)} with 2.3 and 24 GHz carrier frequencies, respectively. As shown in the figures, the channels generated by WiThRay follow the well known fact that the channel with a high carrier frequency is vulnerable to signal blockage and has small coverage. In Fig.~\ref{fig:15}, we show the service areas of four BSs. Each grid is mapped to the BS that provides the strongest received signal power. The figure shows that the serving BS is not always the nearest BS, since some buildings under the BS can block RF signals. These kinds of data generated by WiThRay are useful for practical applications, e.g., exploring the best service position of the BS and designing handover techniques.

WiThRay also follows other important properties of wireless communication channels. In the following subsections, we bring measurements of the channel generated by WiThRay and visualize the channel on the time, frequency, and space domains. Section~\ref{sec4_1} provides the power delay profile (PDP) and shows the relation between delay spread and coherence bandwidth. The channel profile on the frequency domain is analyzed in Section~\ref{sec4_2}. In Section~\ref{sec4_3}, the measurement on the angular domain also shows the spread of CIRs, which determines the beam patterns similar to the array steering vector in \eqref{eq:SteerVec}.

\begin{figure}
	\centering
	\begin{center}$
		\begin{array}{c}
			\subfloat[Experiment scenario.]{\includegraphics[width=0.8\columnwidth]{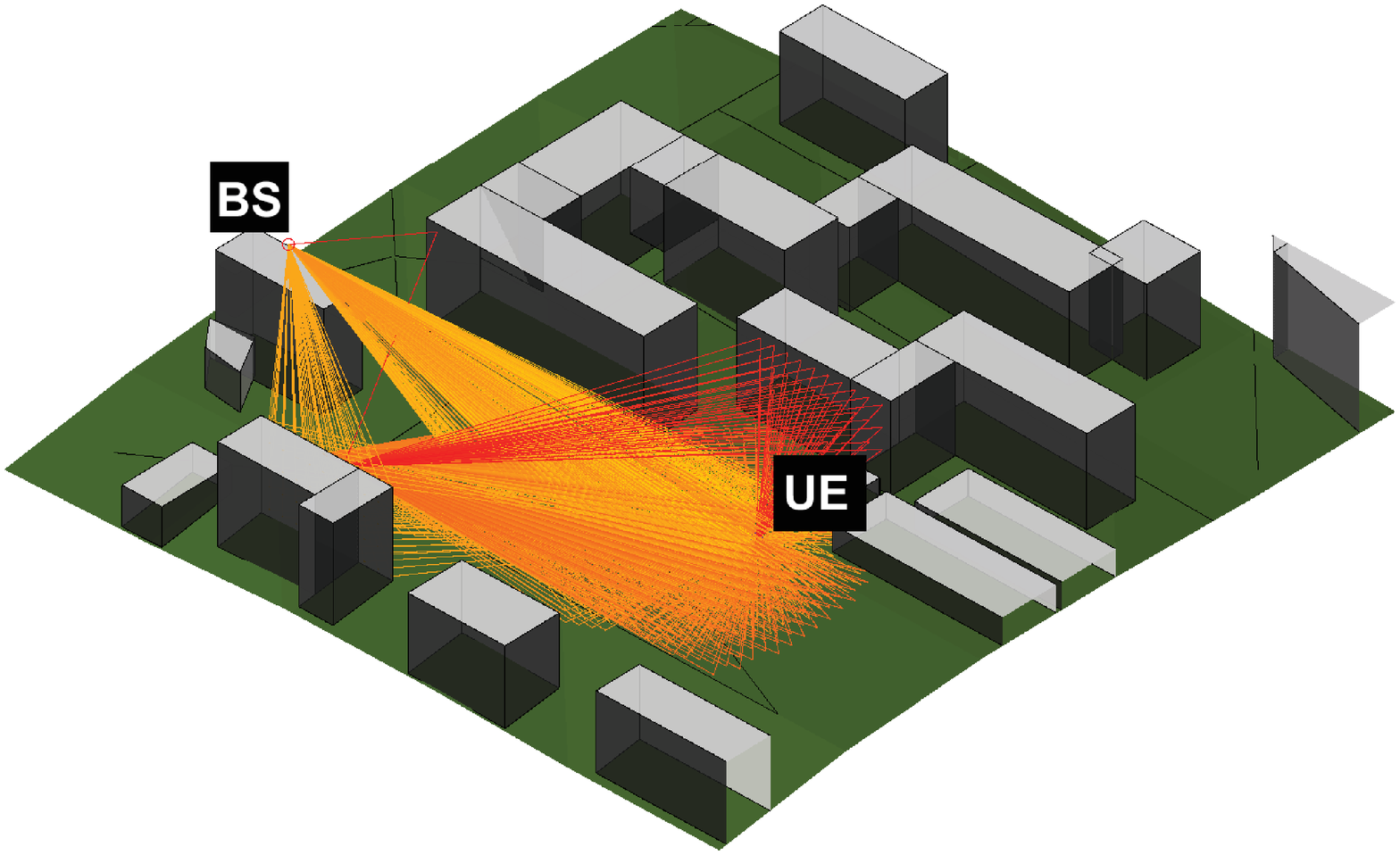} \label{subfig:16-a}}
		\end{array}$\\
	\end{center}
	\begin{center}$
		\begin{array}{ll}
			\subfloat[PDPs with 100 MHz sampling rate and different carrier frequencies.]{\includegraphics[width=0.46\columnwidth]{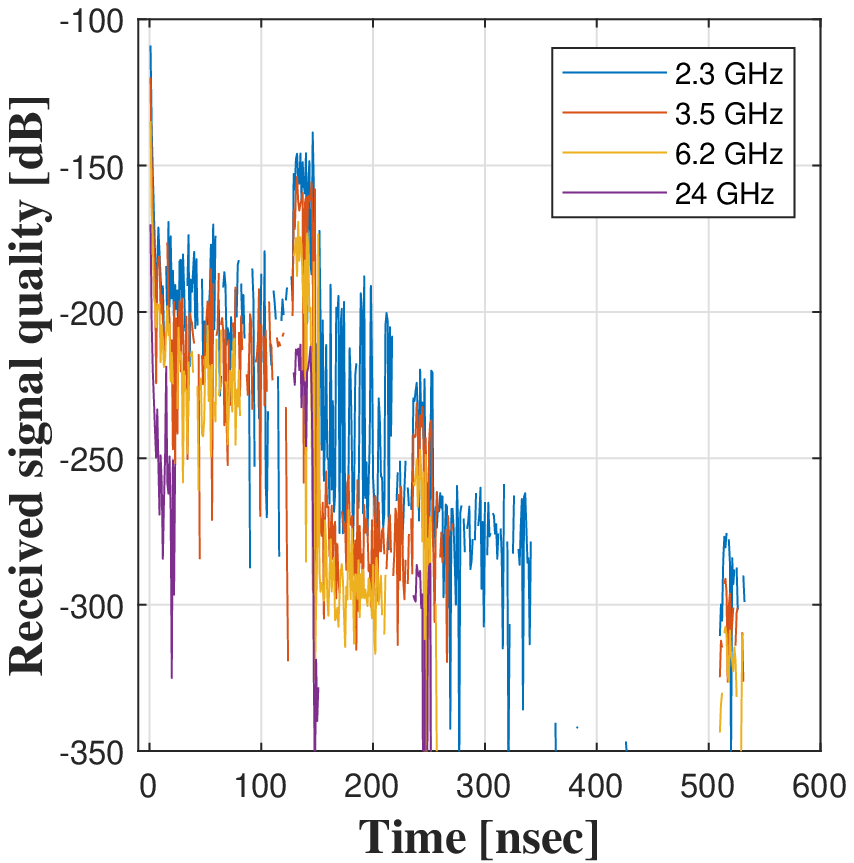} \label{subfig:16-b}} &
			\subfloat[Effective PDPs with 2.3 GHz carrier frequency and different sampling rates.]{\includegraphics[width=0.46\columnwidth]{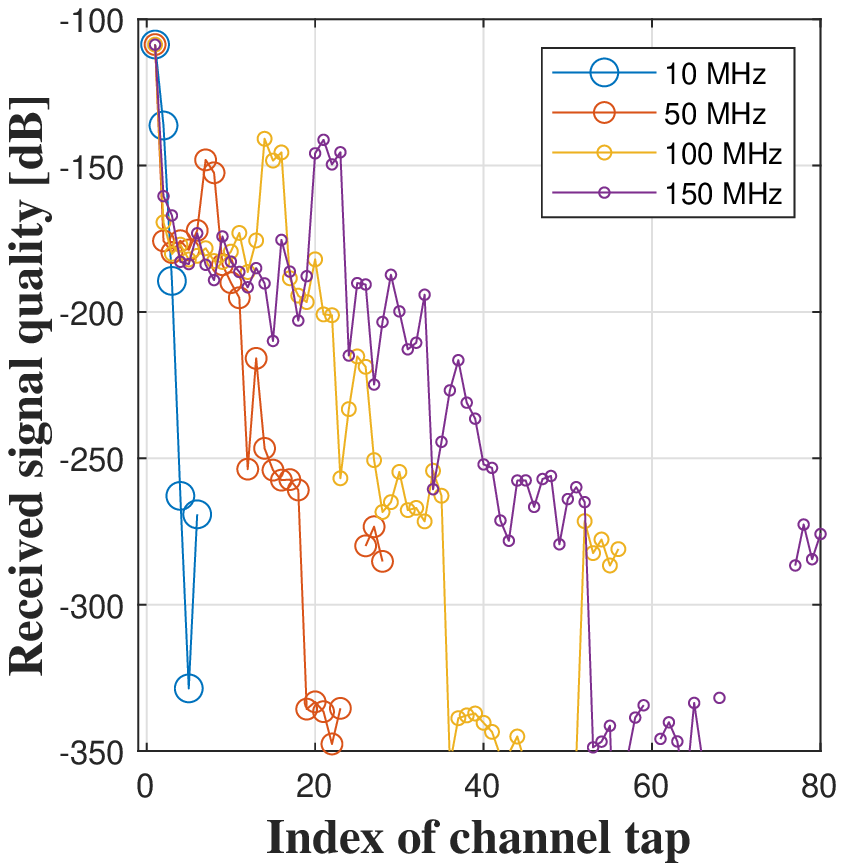} \label{subfig:16-c}}
		\end{array}$
	\end{center}
	\caption{PDP measurement with WiThRay.}
	\label{fig:16}
\end{figure}

\subsection{Power Delay Profile}
\label{sec4_1}
We show the PDP measurement results in Fig.~\ref{fig:16}. The experiment scenario is shown in Fig.~\ref{fig:16}~\subref{subfig:16-a}, where the BS is located on the top of the building, and the mUE approaches to the BS at 1 m/s speed. We test with 100 MHz sampling rate and different carrier frequencies of 2.3, 3.5, 6.2, and 24 GHz as in Fig.~\ref{fig:16}~\subref{subfig:16-b}. The experimental results show that the delay spread decreases as the carrier frequency increases because the pathloss exponent, i.e., the slope of decreasing power in Fig.~\ref{fig:16}~\subref{subfig:16-b}, becomes larger with carrier frequency. The slope in Fig.~\ref{fig:16}~\subref{subfig:16-b} is linear in the log scale and falls exponentially in the linear scale, which follows the Saleh-Valenzuela (SV) model \cite{Saleh:1987,cheng2011new}. With the well-known formula of delay spread\cite{saunders2007antennas}:
\begin{align}
	\tau_s=\frac{\int{\sum_p \tau_p(\mathfrak{t})\left|\alpha_p(\mathfrak{t})\right|^2}d\mathfrak{t}}{\int{\sum_p \left|\alpha_p(\mathfrak{t})\right|^2}d\mathfrak{t}},
\end{align}
the delay spreads of 2.3, 3.5, 6.2, and 24 GHz channels in Fig.~\ref{fig:16}~\subref{subfig:16-b} are 53.3, 48.3, 47.6, and 18.5 nsec, respectively.

\begin{figure}
	\centering
	\includegraphics[width=0.9\columnwidth]{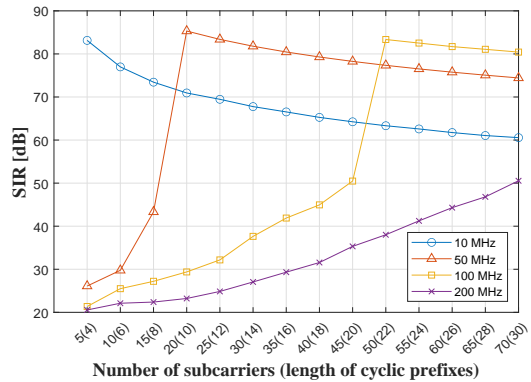}
	\caption{The SIR for mUEs moving at 1 m/s for a system operating at 2.3 GHz carrier frequency.}
	\label{fig:17}
\end{figure}

The measurement results in Fig.~\ref{fig:16}~\subref{subfig:16-b} are the true PDPs, while it is important to investigate the effective PDP that the Rx actually sees. Fig.~\ref{fig:16}~\subref{subfig:16-c} shows the effective PDP with 2.3 GHz carrier frequency, which is heavily affected by the sampling rates, e.g., a small sampling rate combines multiple channel taps into one, making the channel less frequency selective. The multiple CIRs occupy the multiple channel taps for the high sampling frequency, resulting in highly frequency selective channel. The OFDM helps the channel to be flat in a subcarrier. Fig.~\ref{fig:17} depicts the signal-to-interference ratio (SIR) for the 2.3 GHz OFDM channel assuming  equal power allocation across subcarriers. The SIR is then defined as
\begin{align}
	\text{SIR} = \sum_{k=1}^K\frac{|[\boldsymbol{\cH}]_{k,k}|^2}{\sum_{k'\neq k}^K |[\boldsymbol{\cH}]_{k,k'}|^2},
\end{align}
where the OFDM channel $\boldsymbol{\cH}$  is defined in \eqref{eq:OFDMchannel}. The coherence bandwidth of the 2.3 GHz channel with the 82.0 nsec delay spread is 6.1 MHz, which is converted to the number of dominant channel taps for the effective PDP. As shown in Fig.~\ref{fig:16}~\subref{subfig:16-c}, the numbers of dominant channel taps with 10, 50, 100, and 150 MHz sampling rates are 3, 11, 20, and 33, respectively, and the experimental results in Fig.~\ref{fig:17} demonstrate that the SIR becomes extremely large just after using more cyclic prefix than the number of dominant channel taps. Using more subcarriers with fixed total bandwidth, however, makes the bandwidth of each subcarrier smaller and the OFDM symbol length longer. Therefore, due to UE mobility, the channel varies significantly within one OFDM symbol, resulting in fast varying channel taps. The channel $\tilde{\bH}$ in \eqref{eq:OFDMchannel} becomes less circulant for the fast varying channel taps, causing the SIR degradation within the OFDM channel. This is why the 3GPP standard defines flexible OFDM numerology to support high UE mobility with large subcarrier spacing \cite{3gpp2020study}.

\begin{figure}
	\centering
	\begin{center}$
		\begin{array}{c}
			\subfloat[Experiment scenario.]{\includegraphics[width=0.7\columnwidth]{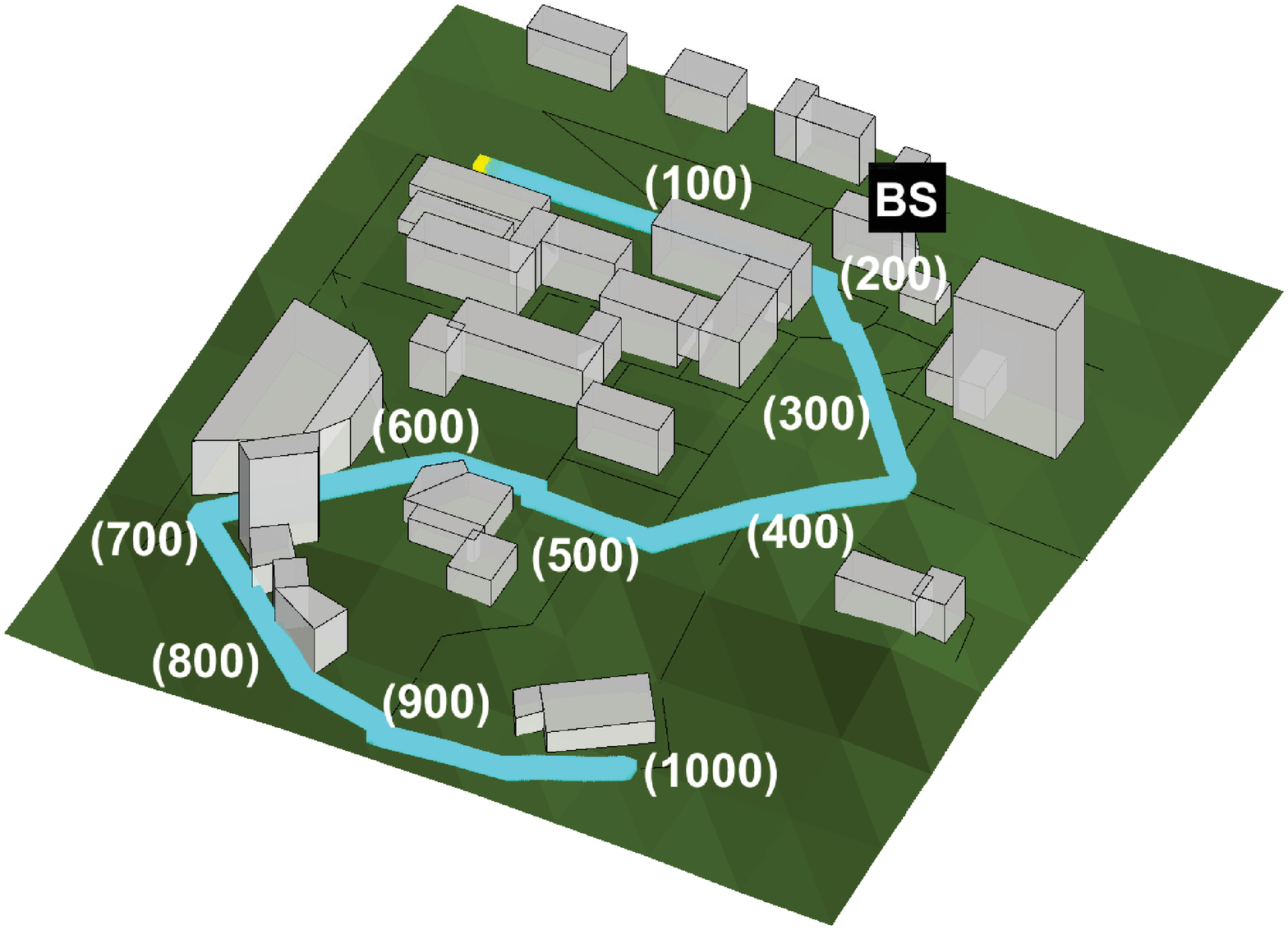} \label{subfig:18-a}}\\
		\end{array}$
	\end{center}
	\begin{center}$
		\begin{array}{cc}
			\subfloat[2.3 GHz carrier frequency.]{\includegraphics[width=0.47\columnwidth]{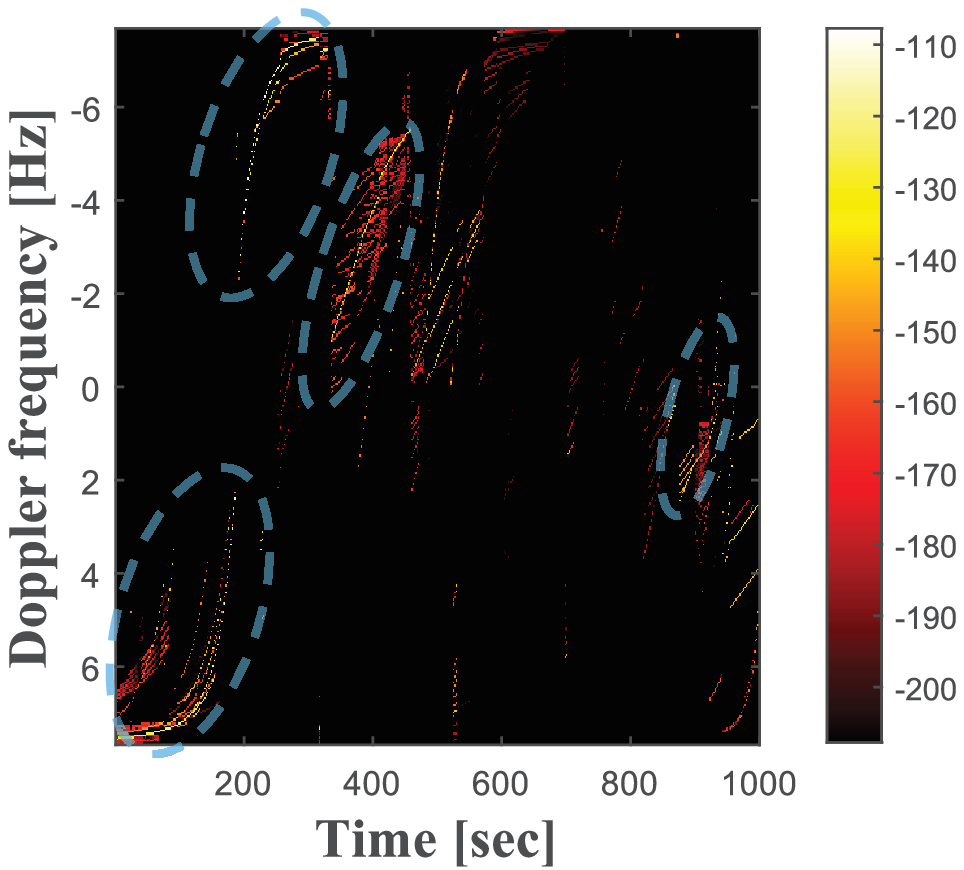} \label{subfig:18-b}}&
			\subfloat[24 GHz carrier frequency.]{\includegraphics[width=0.47\columnwidth]{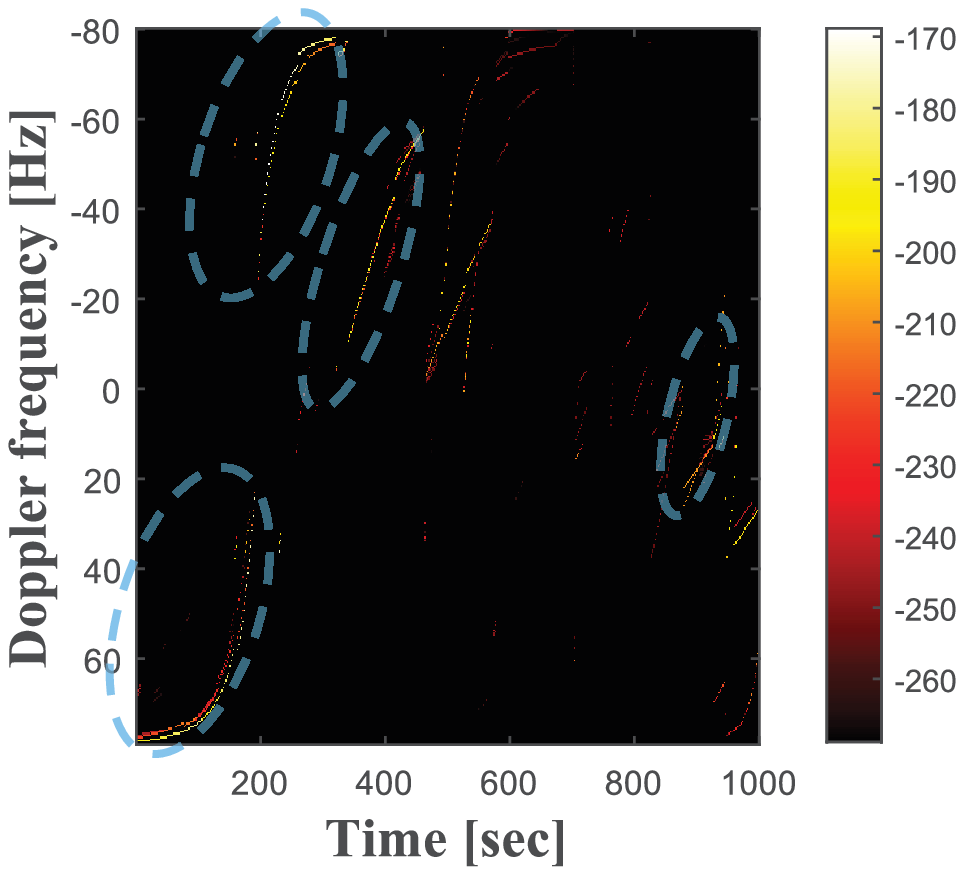} \label{subfig:18-c}}
		\end{array}$
	\end{center}
	\caption{Doppler spread profile measurements for the mUE moving at 1 m/s with WiThRay. The Doppler shift of the LoS path is circled by blue dot line.}
	\label{fig:18}
\end{figure}

\subsection{Doppler Spread Profile}
\label{sec4_2}
The spread of CIRs also appears in the frequency domain. The Doppler shift caused by the mobility of UE changes the carrier frequency. We measure the Doppler spread, specifically, a profile of Doppler shifts, of the mUE moving along the route in Fig.~\ref{fig:18}~\subref{subfig:18-a} and show the results of 2.3 and 24 GHz carrier frequencies in Figs. \ref{fig:18} \subref{subfig:18-b} and \subref{subfig:18-c}, respectively. The received signal power at a certain Doppler frequency is measured by the received signal quality. Since the Doppler shift is a function of the carrier frequency, the range of the Doppler spread is large for the 24 GHz carrier frequency, as shown in Fig.~\ref{fig:18}~\subref{subfig:18-c}. However, Fig.~\ref{fig:18}~\subref{subfig:18-b} shows that the Doppler spread is much prominent for the 2.3 GHz carrier frequency. The time instances are written in Fig.~\ref{fig:18}~\subref{subfig:18-a}, e.g., the mUE approaches the BS during 0-200 sec. As the mUE comes close to the BS, the majority of Doppler shifts have positive values. During 0-122 sec and 221-478 sec, the LoS path exists between the mUE and the BS. The Doppler shift of the LoS path has the dominant received signal quality as sketched by the blue dotted circles in Figs. \ref{fig:18} \subref{subfig:18-b} and \subref{subfig:18-c}.

\begin{figure}
	\centering
	\includegraphics[width=0.9\columnwidth]{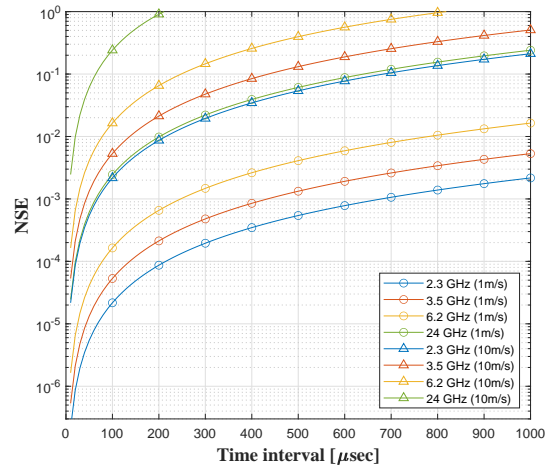}
	\caption{The NSE measurements for 2.3, 3.5, 6.2, and 24 GHz carrier frequencies with the mUE moving at 1 m/s and 10 m/s.}
	\label{fig:19}
\end{figure}

To evaluate the relation between the Doppler spread and the coherence time, we measure the channel variations with 2.3, 3.5, 6.2, and 24 GHz carrier frequencies where the measurement scenario is the same as in Fig.~\ref{fig:16}~\subref{subfig:16-a}. We adopt the normalized square error (NSE) between two time instants $m=m_1$ and $m=m_2$, which is defined as:
\begin{align}
	\text{NSE}[m_1,m_2]=\frac{|\sum_k\mathfrak{h}[m_1,m_1+k]-\mathfrak{h}[m_2,m_2+k]|^2}{|\sum_k\mathfrak{h}[m_1,m_1+k]|^2}.
\end{align}
The discrete channel response $\mathfrak{h}[m,n]$ is defined in \eqref{eq:DiscreteChModel}. The Doppler spread of the channel can be measured as
\begin{align}
	\bar{f}_d=\frac{\int{\sum_p(\bar{f}_p(\mathfrak{t})-\bar{f}_m)^2|\alpha_p(\mathfrak{t})|^2}d\mathfrak{t}}{\int{\sum_p|\alpha_p(\mathfrak{t})|^2}d\mathfrak{t}},
\end{align}
where $\bar{f}_m$ is the mean of the Doppler shifts  that is defined as
\begin{align}
	\bar{f}_m=\frac{\int{\sum_p\bar{f}_p(\mathfrak{t})|\alpha_p(\mathfrak{t})|^2}d\mathfrak{t}}{\int{\sum_p|\alpha_p(\mathfrak{t})|^2}d\mathfrak{t}}.
\end{align}
Since the coherence time is defined as $\tau_c=1/4\bar{f}_d$ in \cite{tse2005fundamentals}, the coherence time of the mUE moving at 1 m/s obtained with the Doppler spread for the 2.3, 3.5, 6.2, and 24 GHz carrier frequencies is 80.73, 46.50, 13.41, and 2.038 msec, respectively. If we define the coherence time as the time that corresponds to $\text{NSE}=10^{-3}$, Fig.~\ref{fig:19} shows that the coherence time obtained with the NSE results, and that with the Doppler spread, match quite well. Fig.~\ref{fig:19} also shows that the mUE moving at 10 m/s velocity has much shorter coherence time than the mUE moving at 1m/s velocity.

\begin{figure}
	\centering
	\begin{center}$
		\begin{array}{c}
			\subfloat[2.3 GHz carrier frequency.]{\includegraphics[width=0.8\columnwidth]{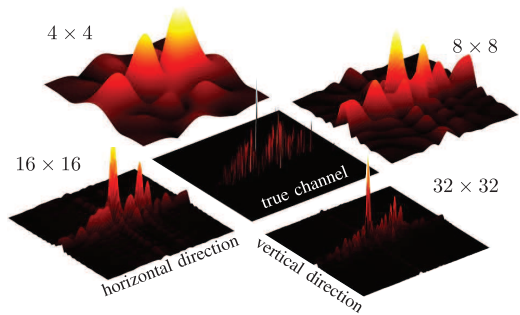} \label{subfig:20-a}}\\
			\subfloat[24 GHz carrier frequency.]{\includegraphics[width=0.8\columnwidth]{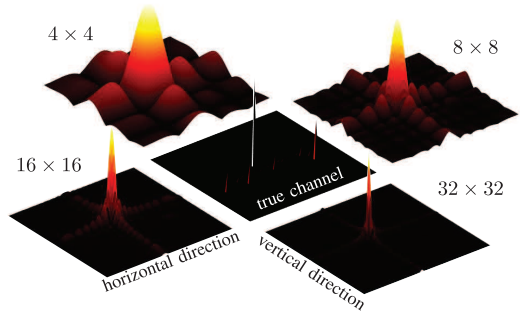} \label{subfig:20-b}}
		\end{array}$
	\end{center}
	\caption{The center image depicts the true angular spread profile on the horizontal and vertical directions, while the other images include the effective angular spread profiles the BS sees with UPA antennas.}
	\label{fig:20}
\end{figure}

\subsection{Angular Spread Profile}
\label{sec4_3}
The angular spread profile provides a measure of the multipath effect in the angular domain, i.e., the horizontal and vertical axes. Similar to the PDP, the angular spread profile is also a function of the carrier frequency. The angular spread profile, however, is heavily affected by the structure of antennas. In Figs. \ref{fig:20} \subref{subfig:20-a} and \subref{subfig:20-b}, we show that the true angular spread profile and the effective angular spread profile that the BS actually sees through UPA antennas for the 2.3 GHz and 24 GHz carrier frequencies, when considering the scenario of Fig.~\ref{fig:16}~\subref{subfig:16-a}. The true angular spread profile of the azimuth $\theta$ and the zenith $\phi$ is defined as follows:
\begin{align}
	\Lambda_\text{true}(\theta,\phi)=\sum_{p\in\cP_t(\theta,\phi)}\int\lvert\alpha_p(\mathfrak{t})\rvert^2 d\mathfrak{t},
\end{align}
with $\cP_t(\theta,\phi)=\{p|\theta_{p,1}^t\in(\theta,\theta+\Delta\theta),\phi_{p,1}^t\in(\phi,\phi+\Delta\phi)\}$. Moreover, the effective angular spread profile for the $N_\text{hor}^t\times N_\text{ver}^t$ UPA is defined as follows:
\begin{align}
	&\Lambda_{\text{eff}}^{N_\text{hor}^t,N_\text{ver}^t}(\theta,\phi)\nonumber \\
	&=\frac{1}{K}\sum_{k=1}^K\left\lvert\sum_{i=1}^{N_\text{hor}^t}\sum_{j=1}^{N_\text{ver}^t}[\boldsymbol{\cH}^{N_\text{ver}^t(i-1)+j,1}]_{k,k}\cdot[\ba(\theta,\phi)]_{N_\text{ver}^t(i-1)+j}\right\rvert^2,
\end{align}
where $[\boldsymbol{\cH}^{\mathfrak{n}_t,\mathfrak{n}_r}]_{k,k}$ is the $k$-th subcarrier channel for the $\mathfrak{n}_t$-th Tx antenna and the $\mathfrak{n}_r$-th Rx antenna, and $[\ba(\theta,\phi)]_{\mathfrak{n}_t}$ is the $\mathfrak{n}_t$-th element of the steering vector in \eqref{eq:SteerVec}.
It is clear from the figures that the 2.3 GHz carrier frequency experiences rich scattering compared to the 24 GHz carrier frequency, while their LoS directions are the same. When the number of antennas is small, e.g., $4\times 4$ UPA, the effective angular spread profiles look quite different from the true angular spread profiles for both carrier frequencies due to insufficient spatial resolution. Therefore, the large number of BS antennas should enable sparsity-based wireless communication techniques.

\section{Example Applications with WiThRay}
\label{sec5}
WiThRay generates channel data satisfying the fundamental properties of wireless channels, as elaborated in Section~\ref{sec4}. The proposed RT-based simulator is versatile enough to evaluate many wireless communication techniques under various environments. This section includes three examples of performance evaluation using WiThRay. We first evaluate uplink channel estimation accuracy for the TV MIMO channel using the OFDM waveform in Section~\ref{sec5_1}. We then compare in Section~\ref{sec5_2} the performance of a multiuser (MU) multiple-input and single-output (MISO) for several techniques: i) matched filtering, ii) zero-forcing (ZF) beamforming, iii) minimum mean squared error (MMSE) beamforming, iv) spatial beamforming, and v) the weighted MMSE (WMMSE) beamforming\cite{Pellaco:2022}. We also provide the simulation results with the RIS for MISO scenario in Section~\ref{sec5_3}.

\begin{figure}
	\centering
	\includegraphics[width=0.7\columnwidth]{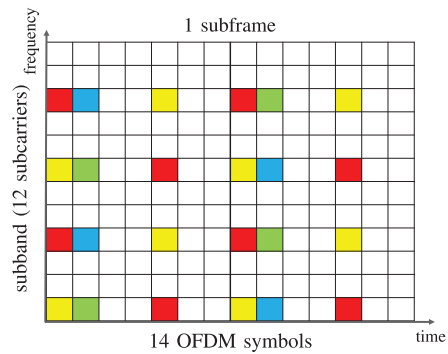}
	\caption{Locations of pilot symbols for the antenna port-1 (red), the antenna port-2 (blue), the antenna port-3 (green), and the antenna port-4 (yellow) in an RB.}
	\label{fig:21}
\end{figure}

\begin{table}
	\caption{The $\text{NSE}_\text{OFDM}$ of the OFDM channel estimation for different subcarrier spacing and speeds of the mUE.}
	\label{table}
	\setlength{\tabcolsep}{3pt}
	\begin{center}
		\begin{tabular}{|l||c|c|c|}
			\hline
			\diagbox{Subcarrier\\spacing}{MT speed}& \textbf{1 m/s} & \textbf{5 m/s} & \textbf{10 m/s} \\ \hline
			\textbf{15 kHz}  & -31.05 dB & -16.94 dB & -10.83 dB\\ \hline
			\textbf{30 kHz}  & -40.08 dB & -25.91 dB & -19.82 dB\\ \hline
			\textbf{60 kHz}  & -48.89 dB & -35.00 dB & -28.90 dB\\
			\hline
		\end{tabular}
	\end{center}
	\label{table:channel_estimation}
\end{table}

\subsection{Application 1: OFDM Uplink Channel Estimation}
\label{sec5_1}
To evaluate the OFDM-based uplink channel estimation, we consider the scenario in Fig.~\ref{fig:16}~\subref{subfig:16-a} with the BS equipped with $2\times 4$ antennas, the UE with a single antenna, the 2.3 GHz carrier frequency, and the 3.84 MHz sampling rate. There are 256, 128, and 64 subcarriers with 15, 30, and 60 kHz subcarrier spacing, respectively, for the experiment. The numerologies with 15, 30, and 60 kHz subcarrier spacing have 12, 6, and 3 subbands, respectively, where each subband consists of 12 subcarriers. There are 14 OFDM symbols in one subframe. A resource block (RB) is shown in Fig.~\ref{fig:21} with the pilot symbol positions for four antenna ports. As derived in Section~\ref{sec4_1}, the coherence bandwidth of the UE moving at 1 m/s is 6.1 MHz, which is much larger than the subband bandwidth values 180, 360, and 620 kHz in this experiment. Therefore, one subband experiences a flat-fading channel.

Table \ref{table:channel_estimation} gives the $\text{NSE}_\text{OFDM}$ of estimated uplink channel when using the uplink pilot symbols in Fig.~\ref{fig:21}. When evaluating the $\text{NSE}_\text{OFDM}$, the subcarrier channel estimated by the pilot symbol in Fig.~\ref{fig:21} is also used as the estimated channel for its nearby resource grids, while more advanced interpolation techniques are possible. The table shows that, as the velocity of mUE increases, the $\text{NSE}_\text{OFDM}$ also increases since the channel varies faster in time. The table also shows that the OFDM channel with the large subcarrier spacing experiences more accurate channel estimation results than the small subcarrier spacing. This result is consistent with Fig.~\ref{fig:19}. As stated in Section~\ref{sec4_1}, for a fixed sampling rate, increasing the subcarrier spacing (i.e., decreasing the number of subcarriers) reduces the channel estimation period, resulting in more accurate channel estimates.

\subsection{Application 2: Downlink MU-MISO Beamforming}
\label{sec5_2}
\begin{figure}
	\centering
	\includegraphics[width=0.9\columnwidth]{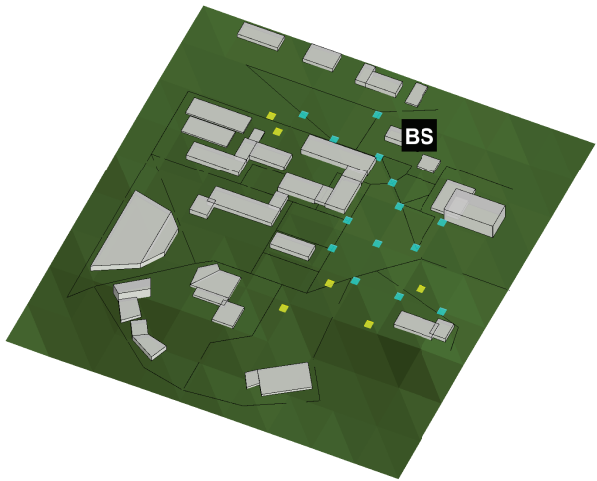}
	\caption{Six UEs colored by yellow are for the first experiment, and totally 20 UEs colored by yellow and cyan are for the second experiment.}
	\label{fig:22}
\end{figure}

In most cases, the performance evaluation for the beamforming employs the sum rate assuming some stochastic channel models, e.g., Rayleigh/Rician fading models or GSCM. However, the sum rate measured in a specific area can be much different from the sum rate obtained from stochastic channel models. To verify how the sum rate changes according to different signal propagation environments, we consider the scenario in Fig.~\ref{fig:22}. There are $K$ randomly located single antenna UEs, and the BS is equipped with $N^t=N_\text{hor}^t\times N_\text{ver}^t$ antennas in the considered scenario.

\begin{figure}
	\centering
	\begin{center}$
		\begin{array}{c}
			\subfloat[Sum rates of six UEs.]{\includegraphics[width=0.95\columnwidth]{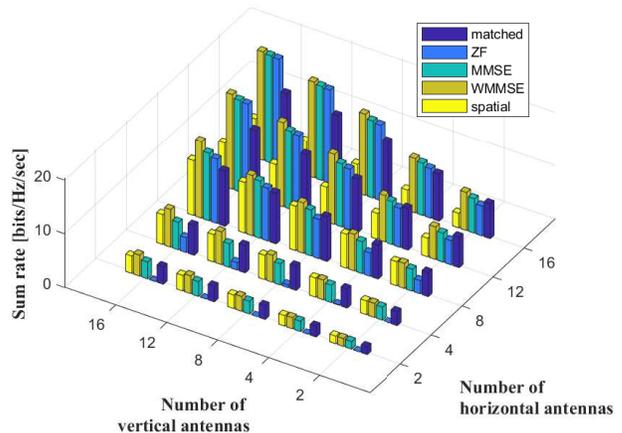} \label{subfig:23-a}}\\
			\subfloat[Sum rates of 20 UEs.]{\includegraphics[width=0.95\columnwidth]{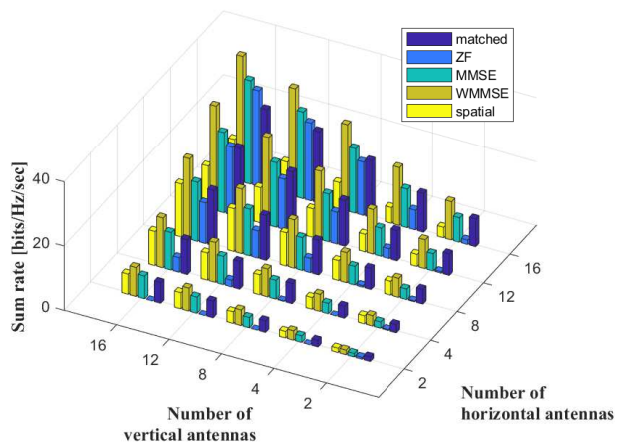} \label{subfig:23-b}}
		\end{array}$
	\end{center}
	\caption{Schematics of the achievable sum rates of six and 20 receivers considering the scenario in Fig.~\ref{fig:22} using matched filter, ZF, MMSE, WMMSE, and spatial beamformers.}
	\label{fig:23}
\end{figure}

Taking only one subcarrier into account, the matched filtering, ZF, MMSE, WMMSE, and spatial beamformers are used to obtain the downlink sum rates. Assuming perfect CSI at the BS, the matched beamformer (before normalization) is represented as
\begin{align}
	\bF_\text{matched} = \bH,
\end{align}
where $\bF_\text{matched}=[\bff_{\text{matched},1}\ \cdots\ \bff_{\text{matched},K}]$ and $\bH=[\bh_1\ \cdots\ \bh_K]$. Without loss of generality, we choose the first subcarrier in this experiment and remove the subcarrier index to simplify the notation. The ZF beamformer $\bF_\text{ZF}$ (before normalization) is obtained as
\begin{align}
	\bF_\text{ZF} = \bH^{-\mathrm{H}}.
\end{align}
The MMSE beamformer is given as
\begin{align}
	\bF_\text{MMSE} = \bH\left(\bH^\mathrm{H}\bH+\frac{N_0W}{P_t}\bI\right)^{-1}.
\end{align}
The WMMSE beamformer is obtained with the iterative algorithm \cite{Pellaco:2022}. To update the beamformer, following variables should be evaluated first:
\begin{align}
	u_k = \frac{\bh_k^\mathrm{H}\bff_{\text{WMMSE},k}}{\sum_{i=1}^K|\bh_k^\mathrm{h}\bff_{\text{WMMSE},k}|^2+N_0 W},\ \forall k\in\cK \label{eq:u_wmmse}\\
	w_k = \frac{\sum_{i=1}^K|\bh_k^\mathrm{H}\bff_{\text{WMMSE},k}|^2+N_0W}{\sum_{i\neq k}|\bh_k^\mathrm{H}\bff_{\text{WMMSE},i}|^2+N_0W},\ \forall k\in\cK. \label{eq:w_wmmse}
\end{align}
With the auxiliary variables $u_k$ and $w_k$, the beamformer is updated as
\begin{align}\label{eq:wmmse}
	\bff_{\text{WMMSE},k} = u_k w_k(\bA+\mu\bI)^{-1}\bh_k,
\end{align}
where $\mu$ is the Lagrange multiplier for the power constraint, and the matrix $\bA$ is defined as
\begin{align}
	\bA=\sum_{i=1}^K w_k|u_k|^2\bh_k\bh_k^\mathrm{H}.
\end{align}
To satisfy the power constraint, the Lagrange multiplier $\mu$ can be obtained by solving a bisection search \cite{Pellaco:2022}. Finally, the optimal WMMSE beamformer $\bF_\text{WMMSE}$ converges as iteratively solving \eqref{eq:u_wmmse}, \eqref{eq:w_wmmse}, and \eqref{eq:wmmse}. 
The spatial beamformer uses the steering vector $\ba(\theta,\phi)$ in \eqref{eq:SteerVec} as
\begin{align}
	\bff_{\text{spatial},k}=\ba(\theta_k,\phi_k),\ k\in\cK,
\end{align}
The beamforming vector $\bff_k$ for the $k$-th UE is normalized as
\begin{align}
	\overline{\bff}_k=\frac{\bff_k}{\lVert\bff_k\rVert},\ k\in\cK.
\end{align}

The sum rate $R$ of the beamformer is obtained as
\begin{align}\label{eq:sum_rate}
	R=\sum_{k=1}^K \log_2\left(1+\frac{P_k|\overline{\bff}_k^\mathrm{H}\bh_k|^2}{\sum_{k'\neq k}P_{k'}|\overline{\bff}_{k'}^\mathrm{H}\bh_k|^2+N_0W}\right),
\end{align}
where $P_k$ is the transmit signal power for the $k$-th UE, $N_0$ is the noise power spectral density, and $W$ is the bandwidth. We consider equal power allocation, i.e., $P_k=P_0,\ \forall k\in\cK$, and the total transmit signal power $P_t=\trace\{\bP\bF^\mathrm{H}\bF\}$ is 10 dBm, where the $K\times K$ matrix $\bP=\diag\{[P_0\ \cdots\ P_0]\}$. We set the carrier frequency as 2.3 GHz and the bandwidth as 50 MHz. Since the noise power spectral density $N_0$ is -174 dBm/Hz, the effective noise power $N_0W$ is -94 dBm.

Fig.~\ref{fig:23} demonstrates the sum rates of the matched filter, ZF, MMSE, WMMSE, and spatial beamformers for the scenario in Fig.~\ref{fig:22} with different numbers of horizontal antennas $N_\text{hor}^t$ and vertical antennas $N_\text{ver}^t$. The WMMSE beamformer consistently provides the best performance. For the case of 20 UEs, its superiority becomes more evident. While the MMSE beamformer is worse than the WMMSE beamformer, it still outperforms the matched filter and ZF beamformer. It is widely known that the ZF beamformer performs well with sufficient antennas \cite{Hoydis:2013}, while the matched filter excels in the opposite case. When the number of antennas is small, the spatial beamformer is comparable to other techniques. While the performance of spatial beamforming keeps increasing as the number of vertical antennas increases, that is not the case as the number of horizontal antennas increases. This is due to the rich scattering in the horizontal direction as shown in Fig.~\ref{fig:20}. With a large number of vertical antennas, e.g., 16, the beamwidth of spatial beamformer becomes quite narrow, making the spatial beamformer fail to cover the large angular spread in the horizontal domain.

\subsection{Application 3: RIS Channel Enhancement}
\label{sec5_3}
\begin{figure}
	\centering
	\begin{center}$
		\begin{array}{c}
			\subfloat[Service area only supported by the RIS.]{\includegraphics[width=0.65\columnwidth]{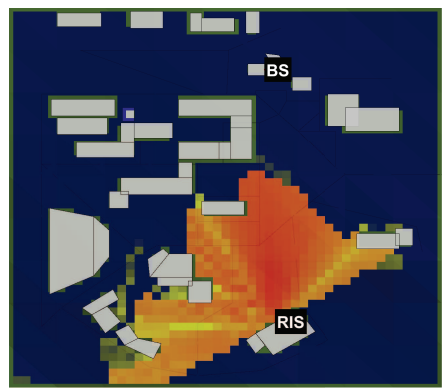} \label{subfig:24_a}}\\
			\subfloat[Service area only supported by the BS.]{\includegraphics[width=0.65\columnwidth]{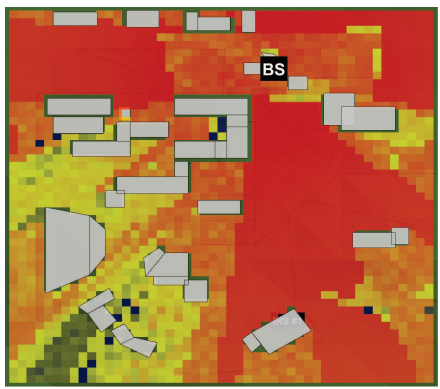} \label{subfig:24_b}} \\
			\subfloat[Service area supported by the BS and RIS.]{\includegraphics[width=0.65\columnwidth]{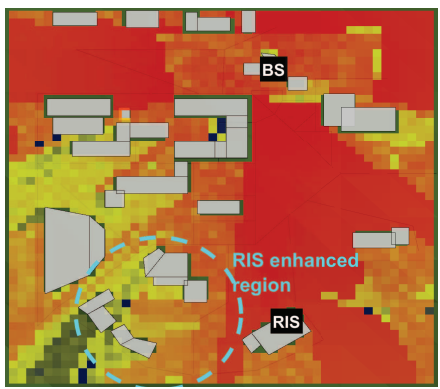} \label{subfig:24_c}}\\
			\subfloat{\includegraphics[width=0.8\columnwidth]{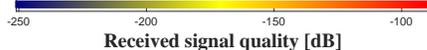} \label{subfig:24_d}}			
		\end{array}$
	\end{center}
	
	\caption{Received signal quality using RIS equipped with 1 m $\times$ 1 m elements. Total size of RIS is 8 m $\times$ 8 m.}
	\label{fig:24}
\end{figure}

The RIS can adjust the direction of reflected signal and enhance the received signal quality at the UE. In this experiment, the UE with a single antenna is served by the BS equipped with $2\times 4$ UPA antennas and the RIS with $8\times 8$ UPA elements as shown in Fig.~\ref{fig:24}. The direct channel from the BS to the UE is $\bh_d\in\mathbb{C}^{8\times1}$, and the channel from the BS to the UE through the RIS is $\bH_i\in\mathbb{C}^{8\times 64}$. Since WiThRay generates the RIS channel as a tensor, the channel from the BS to the RIS and the channel from the RIS to the UE are not separately defined.

The beamforming at the BS and the phase shift at the RIS are jointly adapted in the experiment. For a given phase shift $\bphi$, the beamformer at the BS can be obtained as
\begin{align}\label{eq:matchedbf}
	\bff=\frac{\sqrt{P_t}(\bh_d+\bH_i\bphi)}{\lVert \bh_d+\bH_i\bphi\rVert},
\end{align}
where $\bphi=[\phi_1\ \dots\ \phi_L]$ is the phase shift at the RIS. The matched beamformer in \eqref{eq:matchedbf} is the optimal beamformer for the case of single antenna UE. With the given beamformer $\bff$, the optimal phase shift at the $\ell$-th RIS element is evaluated as \cite{Wu:2019}
\begin{align}
	\phi_\ell=\exp\left(j\angle(\bff^\mathrm{H}\bh_d-\bff^\mathrm{H}[\bH_i]_{:,\ell})\right).
\end{align}
The beamformer and phase shifts are alternatively updated until the gain of effective channel converges.

\begin{figure}
	\centering
	\includegraphics[width=1\columnwidth]{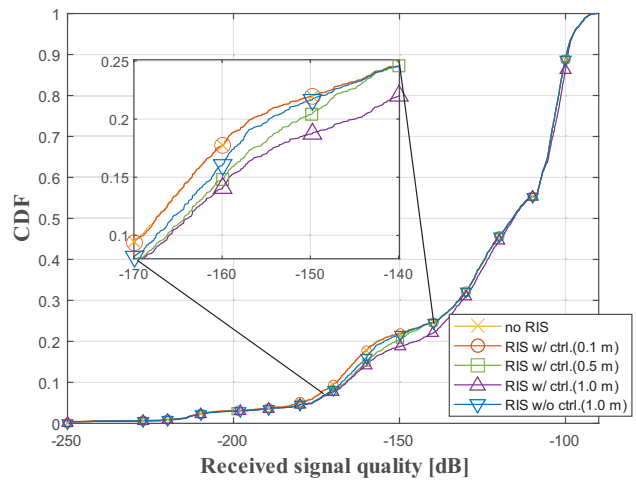}
	\caption{CDF of received signal quality with the BS and RIS using the different size of RIS element.}
	\label{fig:25}
\end{figure}

Fig.~\ref{fig:24}~\subref{subfig:24_a} illustrates the area where signals are received through the RIS with the element size of 1 m $\times$ 1 m. The received signal quality by using the RIS is not comparable to that of the BS, as presented in Fig.~\ref{fig:24}~\subref{subfig:24_b}. Thus, the coverage area enhanced by the RIS is limited to a small region as in Fig.~\ref{fig:24}~\subref{subfig:24_c}. This is reasonable since the RIS will be beneficial only when the signals from the BS are in deep faded. To clearly see the gain of using RIS, we plot the cumulative density function (CDF) of received signal quality with different sizes of RIS element in Fig.~\ref{fig:25}. The RIS gain becomes evident as the element size becomes larger. In the figure, we also plot the performance without controlling the RIS phase shifts. In this case, we set the phase shift $\bphi$ as the all-one vector regardless of the channel. By comparing the RIS with and without control, it is essential to properly control the RIS phase to fully reap off the RIS gain. In general, the results show about 5~10 dB gain by using the RIS for the area where the RIS is helpful.

\section{Conclusions}
\label{sec6}
In this paper, we have introduced WiThRay, an open-source RT-based versatile channel simulator for wireless communications. The channels generated by WiThRay match the fundamental theory of EM wave propagation. This is due to the precise calibration of scattering paths and the adoption of verified EM responses in WiThRay. The proposed BE algorithm adopted in WiThRay accurately and efficiently identifies channel paths, which largely reduces simulation time. WiThRay is designed to enable various experiments with wireless communication technologies over realistic signal propagation environments. 

As representative examples in this paper, we have presented OFDM uplink channel estimation, downlink MU-MISO beamforming, and RIS channel enhancement. The performance evaluation results for the OFDM channel estimation showcased the impact of UE mobility and subcarrier spacing on the channel estimation accuracy. The beamforming application indicated that the WMMSE beamforming shows good performance even with realistic channel data from WiThRay. The matched filtering, ZF, MMSE, and spatial beamformings demonstrate quite good performance, despite their simplicity. The RIS application emphasized the importance of the size of RIS elements and the RIS phase control to fully exploit this futuristic component.

The WiThRay channel simulator can be used for many other wireless communications applications including, but not limited to, sub-6 GHz/mmWave/THz communications, time/frequency/cross division duplexing (TDD/FDD/XDD), and massive MIMO communication systems. WiThRay supports to develop the communication systems with high mobility, e.g., vehicular communications, since the navigating algorithm in WiThRay autonomously creates the trajectories of vehicles. Moreover, WiThRay supports the emerging technology of RISs, where the EM response on the RIS panel is precisely modeled. This versatile simulator WiThRay will provide insights into the future of 6G communication.

\end{document}